\documentclass[%
preprint,
 amsmath,amssymb,
 aps,
prb,
floatfix,
]{revtex4-2}

\usepackage{dcolumn}% Align table columns on decimal point
\usepackage{bm}% bold math
\usepackage{amsmath} %for tagging equations
\usepackage{graphicx} % Include figure files
\usepackage{array} % For stretching table row
\usepackage{titlesec}
\usepackage[hidelinks]{hyperref} % add hypertext capabilities
\usepackage{color}
\usepackage{float}
\usepackage[dvipsnames]{xcolor}
\usepackage{physics}
\usepackage{soul}
\usepackage{ulem}
\usepackage{subcaption}
\usepackage{multirow}
\usepackage[font=small]{caption}
\usepackage{upgreek}
\usepackage{makecell}
\usepackage{tabularx}
\usepackage{natbib}
\usepackage[hidelinks]{hyperref} % add hypertext capabilities

\renewcommand{\thesection}{\arabic{section}}

\makeatletter
\usepackage{titlesec}
\titleformat{\chapter}
  {\normalfont\LARGE\bfseries\color{darkblue}}{\thechapter.}{1em}{}
\newcommand{\beginsupplement}{%
        \setcounter{table}{0}
        \renewcommand{\thetable}{S\arabic{table}}%
        \setcounter{figure}{0}
        \renewcommand{\thefigure}{S\arabic{figure}}%
        \setcounter{section}{0}
        \renewcommand{\thesection}{S\arabic{section}}
     }

\begin{document}

\title{Einstein-de Haas torque as a discrete spectroscopic probe allows nanomechanical measurement of a magnetic resonance}
\author{K.R. Fast$^{1,3,^{\ast}}$, J.E. Losby$^{2,3,^{\ast}}$, G. Hajisalem$^{2,3}$, P.E. Barclay$^{2,3}$, M.R. Freeman$^{1,3}$}

\affiliation{1. 
Department of Physics, University of Alberta, Edmonton, Alberta T6G 2E1, Canada%\\
}
\affiliation{2. 
Department of Physics and Astronomy, University of Calgary, Calgary, Alberta T2N 1N4, Canada%\\
}
\affiliation{3.
Nanotechnology Research Centre, National Research Council
of Canada, Edmonton, Alberta T6G 2M9, Canada%\\
}

\email{Both authors contributed equally to this work. email: pbarclay@ucalgary.ca , freemanm@ualberta.ca}

\date{\today}
\begin{abstract}
The Einstein-de Haas (EdH) effect is a fundamental, mechanical consequence of any temporal change of magnetism in an object.  EdH torque results from conserving the object's total angular momentum: the angular momenta of all the specimen's magnetic moments, together with its mechanical angular momentum.  Although the EdH effect is usually small and difficult to observe, it increases in magnitude with detection frequency.  We explore the frequency-dependence of EdH torque for a thin film permalloy microstructure by employing a ladder of flexural beam modes (with five distinct resonance frequencies spanning from 3 to 208 MHz) within a nanocavity optomechanical torque sensor via magnetic hysteresis curves measured at mechanical resonances.  At low DC fields the gyrotropic resonance of a magnetic vortex spin texture overlaps the 208 MHz mechanical mode.  The massive EdH mechanical torques arising from this co-resonance yield a fingerprint of vortex core pinning and depinning in the sample.  The experimental results are discussed in relation to mechanical torques predicted from both macrospin (at high DC magnetic field) and finite-difference solutions to the Landau-Lifshitz-Gilbert (LLG) equation.  A global fit of the LLG solutions to the frequency-dependent data reveals a statistically significant discrepancy between the experimentally observed and simulated torque phase behaviours at spin texture transitions that can be reduced through the addition of a time constant to the conversion between magnetic cross-product torque and mechanical torque, constrained by experiment to be in the range of 0.5 - 4 ns.
%Valid PACS numbers may be entered using the \verb+\pacs{#1}+ command.
\end{abstract}
\maketitle

\section{Introduction}
The Einstein-de Haas (EdH) effect is the result of a landmark gyromagnetic investigation conducted just over a century ago that established the fundamental connection between magnetism and mechanical angular momentum \cite{Einstein1915}.  In the EdH experiment, mechanical rotation of a suspended ferromagnetic body was observed as it underwent an alternating change in net magnetization at the torsional resonance frequency. The effect was very small and, in the first observation, subject to systematic experimental uncertainty comparable in magnitude to the measured torque.  Regrowing interest in the EdH effect \cite{Wallis2006,Jaafar2009,Chudnovsky2014,Zarzuela2015,Wells2019,Ruckriegel2020,Mori2020,Garanin2021,Dednam2022} has emerged with the miniaturization of torque sensors through micro- and nano-fabrication \cite{Brooks1987,Cleland1996,Carr1997,Alzetta1999,Moreland2001,Jander2001,Chudnovsky2014}.  Small mechanical sensors couple well to magnetic torques generated in small magnetic specimens affixed to them \cite {Losby2018}.  Micromechanical EdH investigations by Wallis et al. determined the magnetomechanical ratio (the inverse of the gyromagnetic ratio) of a thin film of permalloy deposited on a silicon nitride cantilever \cite{Wallis2006}. In another investigation, spin wave currents induced through the spin-Seebeck effect transferred angular momentum to a mechanical resonator created in bulk yttrium iron garnet \cite{Harii2019}.  In these cases, the mechanical resonance frequencies of the sensors were on the order of tens of kilohertz. Advances in nanomechanical sensing, and particularly the incorporation of cavity optomechanical readouts \cite{Eichenfield2009,Kim2013,Aspelmeyer2014,Forstner2014,Li2020}, have introduced the displacement sensitivity needed for mechanical probes of physical phenomena to progress to radio and microwave frequencies, overcoming the challenge from amplitudes of driven motion generally decreasing as oscillation frequencies increase.  

Measurements to higher driving frequencies expand the scope of what can be learned about magnetism through mechanical sensing of torque.  EdH torques grow with frequency, and probe AC magnetic susceptibilities via the intrinsic angular momentum embedded and locked to the magnetization.  At high-enough frequencies, spin resonances should be observable directly through mechanical torque, and information may emerge about the conversion from magnetic to mechanical drive torque.

Mechanically-sensed, AC magnetic field-driven cross product torque is a probe of magnetic anisotropy.  The AC cross product torques are frequency-independent when the system remains close to magnetic equilibrium -- that is, without thermally-activated, slow dynamics and far enough from spin resonances.  The AC EdH torques, in contrast, increase linearly with frequency in the same regime.  Nanomechanical EdH measurements at 2.7 MHz demonstrated that the effect can be comparable to or even larger in magnitude than conventional cross product magnetic torques in nearly-demagnetized yttrium iron garnet microdisks at low DC applied fields, and confirmed the 90 degree phase difference expected between these two sources of torque when their respective driving fields are in phase \cite{Mori2020}.  Even for magnetically-saturated specimens, the EdH component in an AC torque study is expected to grow from negligible at kHz frequencies into the dominant source of torque when a spin resonance overlaps the mechanical detection frequency and where the resonant enhancement of AC magnetization enables the direct mechanical detection of nonequilibrium magnetism. Torque magnetometry performed at such frequencies could enable coherent coupling of mechanics and spin, as well as open potential pathways for mechanical control of magnetism \cite{Kovalev2005, Kovalev2006}.  

\subsection{Multi-modal torque sensor}
\begin{figure*}[ht] 
	\centering
    \includegraphics[width=\linewidth] {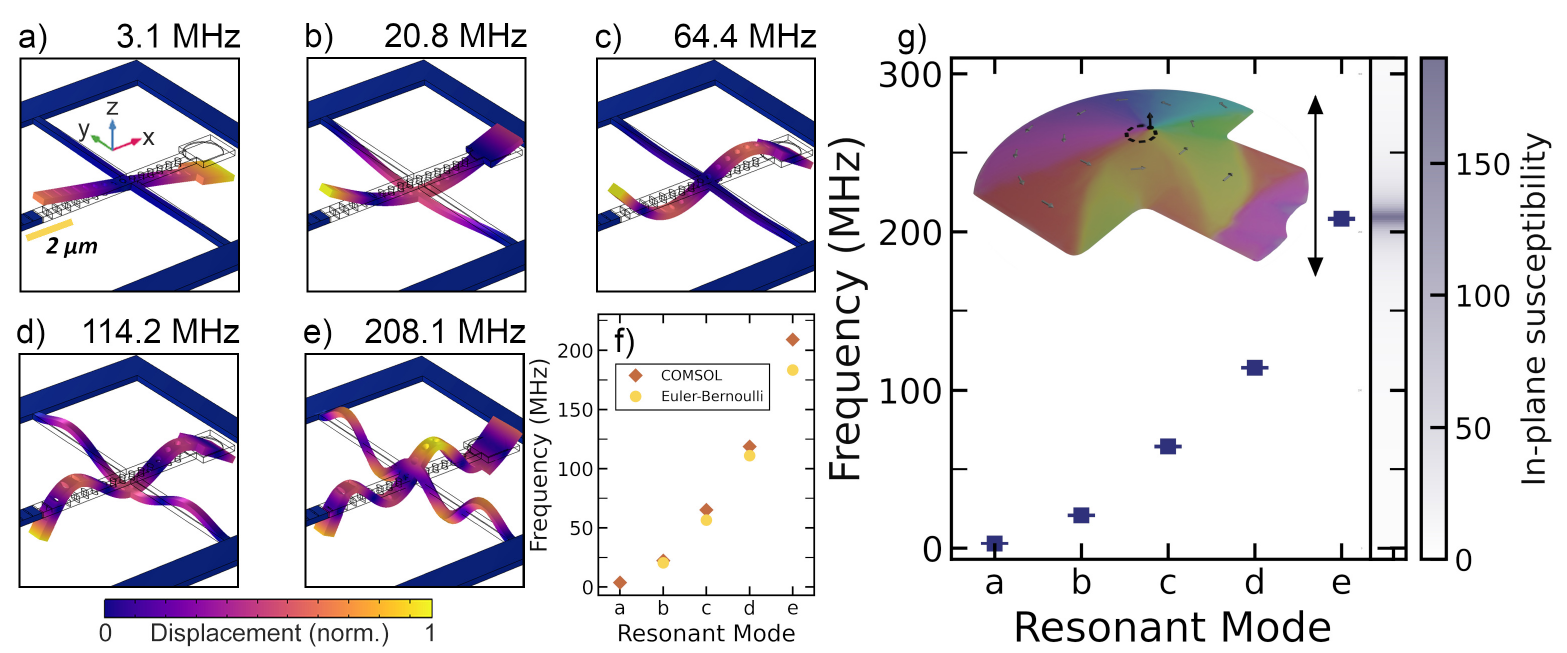}
	\caption{\textbf{Mechanical displacement profiles of the multimodal RF mechanical torque sensor.}   (a) The fundamental torsional mode. A thin-film island of permalloy (Ni$_{80}$Fe$_{20}$) is deposited on the paddle of the of the free end (outlined on the right side of the stationary resonator wire-frame view). (b)-(e) Higher order free-free beam mode profiles.  (f) Mode frequencies of (a)-(e) computed by finite element analysis in COMSOL Multiphysics \cite{COMSOL} (filled diamonds) and predicted by Euler-Bernoulli theory assuming free-free boundary conditions (filled circles). (g) Mode frequencies from experiment; error bars indicate resonance linewidth. Simulation of dimensionless magnetic susceptibility over the experimental frequency range at a DC bias field of 2.39 kA/m is indicated by the colour bar on the right axis. A magnetic resonance (as illustrated by the cartoon in the upper panel where an RF-field driven magnetic vortex orbits around its equilibrium position) overlaps the 208 MHz mechanical mode. Simulated magnetic resonance frequencies for bias fields between 0 and 15 kA/m are indicated by the black arrow. }  
	\label{fig:mechmodes}
\end{figure*}

Continuously-swept frequency measurements would require torque sensitivity away from mechanical resonances at a level that has not yet been demonstrated.  Alternatively, observations at the discrete frequencies within a ladder of resonance modes could permit coresonant detection of mechanical and spin resonances through higher order mechanical modes, as well as utilizing the benefit of EdH torque scaling with frequency.  The nanocavity optomechanical torque sensor used in \cite{Wu2017,Hajisalem2019} is suited to such a study.  Our torque sensor geometry and the spatial displacement profiles of the mechanical mode ladder used for torque measurements are shown in \autoref{fig:mechmodes}.  At the heart of the sensor is a nanomechanical silicon beam along the $x$-direction, with a paddle at one end to host the magnetic specimen -- here, a mushroom-shaped thin-film permalloy (Ni$_{80}$Fe$_{20}$) island.  The sensor beam is suspended from its middle by a symmetrical pair of silicon torsion bars along the $y$-direction.  This central mounting of the beam is a key to observing a series of mechanical modes at different frequencies, all driven by RF magnetic $y$-torques from the sample.  The other crucial feature is the optical nanocavity at the other end of the sensor beam, to detect the ensuing mechanical oscillations.  1D photonic crystal mirrors are embedded in a split-beam geometry and create an optical cavity with maximum field magnitude in the gap between the sensor and stationary beams \cite{Hryciw2013}. 

The lowest-frequency mechanical resonance mode is the fundamental torsional mode shown in \autoref{fig:mechmodes}a.  The mode ladder beyond the fundamental torsion resonance exploits the similarity of the suspended sensor's mechanical modes to those of a beam with free ends, thanks to cooperative flexing of the side supports.  These modes are all responsive to to $y$-directed RF torque, arising both from the cross products between $H_z^{\mathrm{RF}}$ fields and $x$-directed net magnetic moment, as well as from $H_y^{\mathrm{RF}}$-driven EdH torques.  The optical nanocavity has sufficient displacement sensitivity to detect the first four modes in the ladder, at 21, 64, 114, and 208 MHz; the corresponding eigenmodes found by finite-element simulation with COMSOL Multiphysics \cite{COMSOL} are shown in \autoref{fig:mechmodes}b - \ref{fig:mechmodes}e.  Mode frequencies predicted by the Euler-Bernoulli approximation for a thin beam calculated using the material and structural parameters of the sensor (Supplementary \autoref{suppSect:EBFrequencies} \cite{Supplement}) are also displayed in \autoref{fig:mechmodes}f alongside the experimentally-measured (Supplementary \autoref{suppSect:MechFrequencies} \cite{Supplement}) and COMSOL-estimated eigenmode frequencies.  The Euler-Bernoulli formula yields rough agreement with measured mode frequencies, with higher discrepancies when the support beam undergoes more twisting and there is a node in the sensor beam displacement profile at the intersection point (\autoref{fig:mechmodes}c,e) that is unaccounted for in the Euler-Bernoulli boundary conditions.

\subsection{Basic expectations for frequency- and field-dependencies of EdH and cross product torques}
Contemporary studies of EdH mechanical torques have been restricted to fully- or partially-demagnetized specimens, which are driven around minor hysteresis loops by an RF field to yield an oscillating magnitude of net magnetic moment.  In addition, the drive frequencies have been low enough to assume that the magnetization remains in \cite{Wallis2006,Lim2014,Mori2020} or close to \cite{Mori2020} (i.e. slow dynamics assisted by thermal activation) equilibrium with the applied DC and RF magnetic fields. To gain insight into the conventional and EdH torque contributions under arbitrary drive conditions, we developed and applied a macrospin model to solve the Landau-Lifshitz-Gilbert (LLG) equation for torques in the limit where an almost uniformly magnetized specimen can be approximated by a single, giant spin subject to the sample's magnetic anisotropies (See Supplementary \autoref{suppSect:Macrospin} \cite{Supplement}).  The macrospin code executes hundreds of times faster than a corresponding finite-difference micromagnetic LLG simulation.  Solutions for RF drives are passed through a  software lock-in emulator to extract torque magnitudes and phases (the same format as the experimental data; details in Supplementary \autoref{suppSect:LockIn} \cite{Supplement}).
%The anisotropy driven torque, $\tau^{\mathrm{anis}} = M_{\mathrm{s}}V(\mathbf{m} \times \mu_0\mathbf{H}_{\mathrm{anis}})$ is the torque driven by internal anisotropy field. This field is defined within the macrospin framework as 

Mechanical reaction torques consist of the external field contributions from the full solutions to the LLG equation plus the Einstein-de Haas, angular momentum-conserving contributions.  For a simulation set up to mimic our experiment, we focus on the torque component in the plane of the thin film and perpendicular to the $x$-directed DC field ($\hat{y}$).  This component ($y$-torque in our coordinate system) drives all five of the mechanical modes used in the measurements reported here.  The resultant $y$-torques are computed for two directions of AC field drive: along $\hat{z}$ in the traditional cross product $y$-torque configuration, and along $\hat{y}$ in the EdH configuration.

\begin{figure}[ht] 
	\centering
    \includegraphics[width=0.7\linewidth] {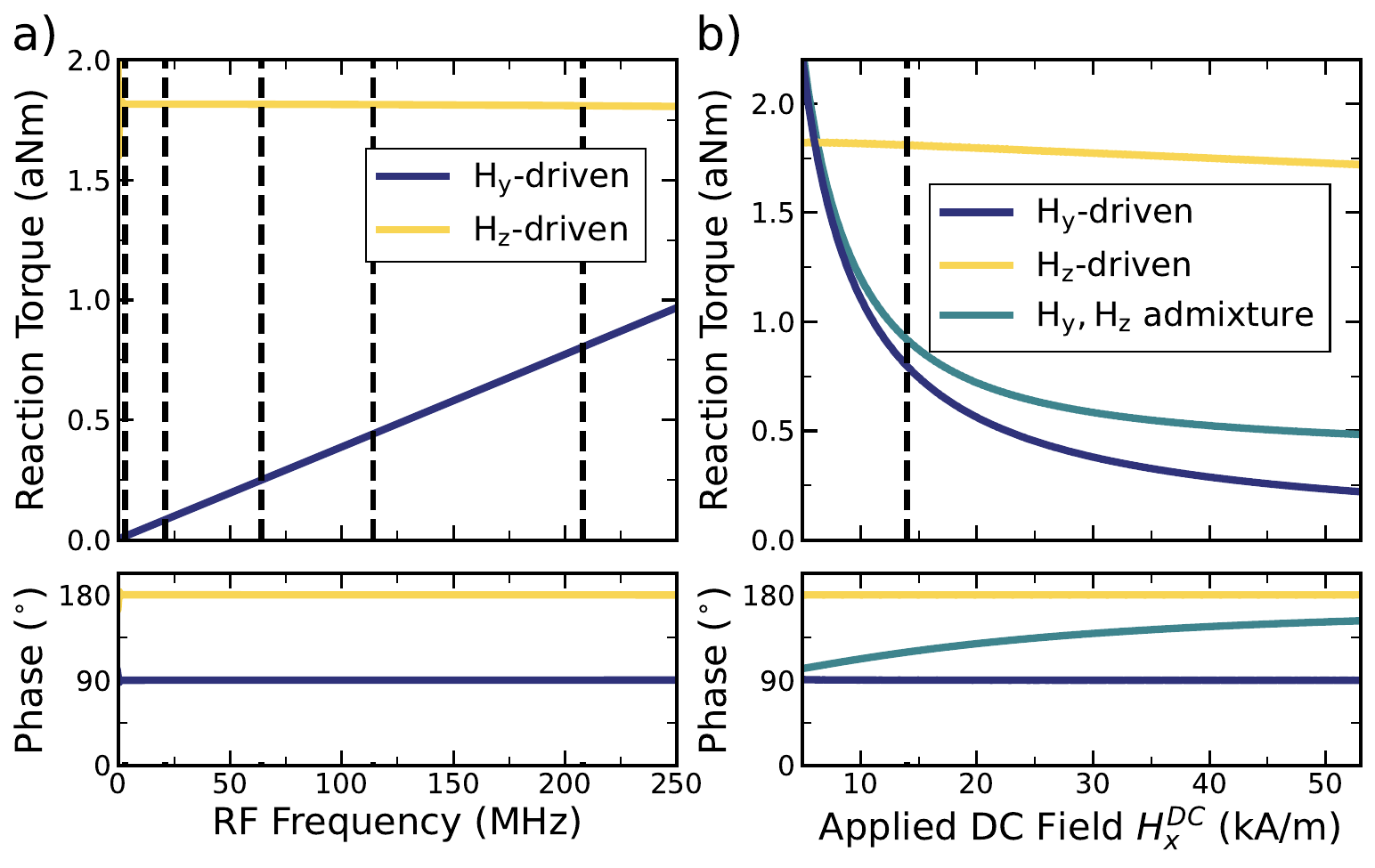}
	\caption{\textbf{Macrospin simulation results.} Frequency (a) and DC field (b) dependencies of simulated macrospin reaction torques for a 470 nm diameter permalloy sphere with uniaxial hard axis anisotropy along $\hat{z}$, uniaxial anisotropy constant $K_u = 340$ kJ/m$^3$, saturation magnetization $M_s$=767 kA/m and Gilbert damping $\alpha$=0.01. The dashed lines in (a) indicate experimentally measured frequencies of the torque sensor used in this work (3.1, 20.8, 64.4, 114.2, and 208.1 MHz). The DC field of 14 kA/m used in (a) is indicated by a dashed line in (b). The DC field sweeps in (b) are calculated for a drive frequency of 208 MHz. Admixture of $H_y$- and $H_z$-driven torques are shown in teal, for a $H_y^{\mathrm{RF}}$ amplitude of 50 A/m and a $H_z^{\mathrm{RF}}$ amplitude of 12.5 A/m. The resultant phase of the admixture-driven torque has a significant dependence on the applied DC field, in contrast to the constant phases of the the purely $H_y$- and $H_z$-driven torques.
}  
	\label{fig:macrospin}
\end{figure}

Predicted resultant torque magnitudes and phases for two different cuts through experimental parameter space are shown in \autoref{fig:macrospin} for a spherical macrospin having the same total moment as the permalloy island and with a uniaxial hard axis along $\hat{z}$. The anisotropy energy density equivalent to the shape anisotropy of the microstructure, which has demagnetizing factors $N_x$=0.042, $N_y$=0.045, and $N_z$=0.913, is -340 kJ/m$^3$. A 50 A/m AC field is applied along $\hat{z}$ ($H_z$) or $\hat{y}$ ($H_y$) to promote conventional or EdH torques about the $\hat{y}$ axis.  As seen in \autoref{fig:macrospin}a, at frequencies in the range of 100 MHz the EdH torque generated by $H_y$ is not a small effect, it is in the same order of magnitude as the cross product torque.  This is noteworthy because for these parameters, where the DC applied field is small compared to the effective field from the uniaxial anisotropy, the resultant torque generated by $H_z$ is very close to the maximum value possible for that amplitude of drive.  The $H_y$-driven torque is also phase shifted by $90^{\circ}$ relative to the drive phase. 

As a function of increasing DC field strength (\autoref{fig:macrospin}b), the $H_z$-driven cross product torque decreases slowly as the angular excursion of the moment is reduced by the DC field below the limit set at zero field by the magnetic anisotropy.  The $H_y$-driven EdH torque decreases more rapidly with DC field because the angular excursion of the moment in the easy plane is set by $H_x$. An admixture of $H_y$- and $H_z$-driven torques similar to experimental conditions produces torques which exhibit qualitative similarity (in both magnitude and phase) to measured torques in the quasi-saturated spin texture (\autoref{fig:hysteresis}c).

\section{Magnetic torque observations}
A DC bias field in the x-direction is provided by a permanent magnet mounted on a linear translation stage. Hysteresis measurements are acquired by sweeping the permanent magnet. The thin film permalloy micromagnet self-demagnetizes in low field and therefore it is sufficient to measure unipolar hysteresis loops.  

\begin{figure}[ht]
    \centering
    \includegraphics[width=0.7\linewidth]{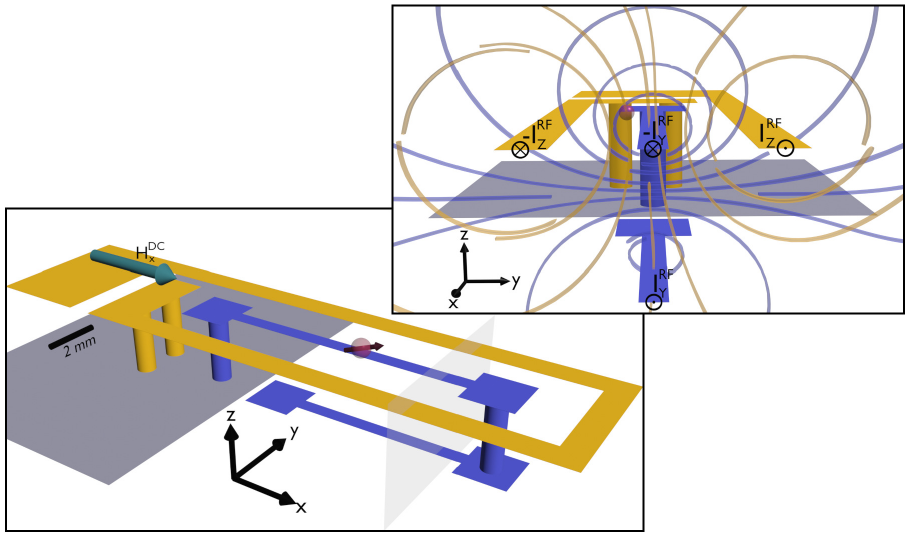}
    \caption{\textbf{Strip line field coil geometry.} The design of the transmission lines providing RF fields to the sample is shown. Two separate coil loops are present on a single circuit board. Input and return currents are indicated in the upper figure as $\pm I_z^{\mathrm{RF}}$ (yellow coil) and $\pm I_y^{\mathrm{RF}}$ (purple coil). The ground plane is shown in gray. Magnetic field lines from each coil (calculated with COMSOL Multiphysics \cite{COMSOL}) are shown in the top panel and illustrate the admixtures of $z$- and $y-$drive field components at the sample location (indicated by the red sphere).}
    \label{fig:fieldGeometry}
\end{figure}

For transduction, a dimple created at the center of a tapered section of optical fiber is placed on the stationary beam mirror to couple light into the photonic crystal cavity.  The fiber-coupled laser wavelength is tuned to a position of maximum slope at the cavity resonance in the fiber's transmission spectrum, such that the perturbations of the optical field from beam vibrations are encoded in the intensity and  can be read out using lock-in detection \cite{Wu2017, Hajisalem2019}. We apply magnetic torques to the permalloy island using a custom built, multi-layer circuit board (\autoref{fig:fieldGeometry}) with single-turn transmission line coils for generating RF magnetic fields in the $\hat{y}$ and $\hat{z}$ directions.  Respectively, an RF current $I_y^{\mathrm{RF}}$ ($I_z^{\mathrm{RF}}$) applied to the vertical coil (horizontal coil) yields a predominantly $H_y^{\mathrm{RF}}$ ($H_z^{\mathrm{RF}}$) magnetic drive field at the sample location, with a small admixture of $H_z^{\mathrm{RF}}$ ($H_y^{\mathrm{RF}}$).  The present experimental setup does \emph{not} have a means for determining the absolute drive field phases at the sample position.  We confirmed through sniffer coil measurements with lock-in detection that the $H_z^{\mathrm{RF}}$ and $H_y^{\mathrm{RF}}$ drive fields are accurately in phase with one another at the sample, and the experiment therefore has good sensitivity to changes in the relative phases of the EdH and cross product torques (Supplementary \autoref{suppSect:SnifferCoil} \cite{Supplement}).

Some parasitic coupling between the RF drives and the photoreceiver exists in the setup and is characterized by companion measurements acquired with the laser tuned away from the optical cavity resonance, to minimize the mechanical signal.  The mechanically-transduced signal magnitudes and phases are then obtained by phasor subtracting the tuned-away backgrounds from the signals at optimal optomechanical coupling (see Supplementary \autoref{suppSect:PhasorSubs} \cite{Supplement}).

\subsection{Measurements at 3, 21, 64, and 114 MHz modes: frequency-dependence of EdH torque}

\begin{figure}[ht] 
	\centering
    \includegraphics[width=\linewidth] {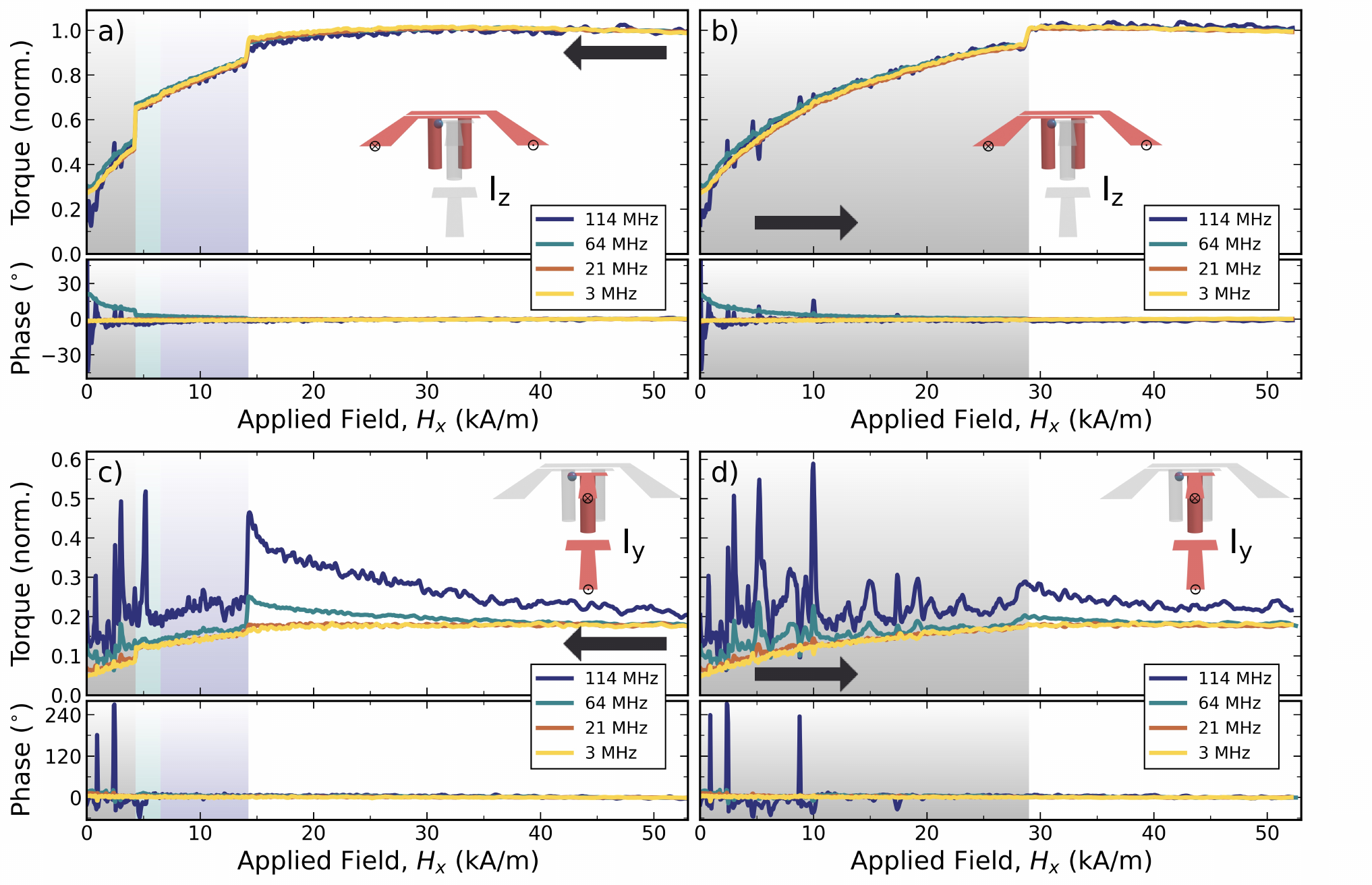}
	\caption{\textbf{Torque measurements of hysteresis at 3, 21, 64, and 114 MHz for two orientations of RF drive field.} High to low field (a,c) and reverse direction (b,d) hysteresis for the $I_z^{\mathrm{RF}}$- (a,b) and $I_y^{\mathrm{RF}}$- (c,d) driven modes up to 114 MHz, with corresponding phase signals in the lower panel of each quadrant. Spin textures present throughout the field sweeps are highlighted with different background shading.  The unshaded section is the quasi-uniform magnetization texture, the purple is the nucleation of vortex core topologies near the edge of the disk, the blue is the single-vortex nucleation and transition region, and the gray is a single-vortex after relaxation of magnetization in another part of the sample. The drive coil used in each case is highlighted in red in the corresponding inset schematic.}  
	\label{fig:hysteresis}
\end{figure}

Normalized hysteresis loops initialized at high field for the 3, 21, 64, and 114 MHz mechanical modes are shown in \autoref{fig:hysteresis}. The normalization method takes advantage of the drive field admixtures and is described in detail in Supplementary \autoref{suppSect:Normalization} \cite{Supplement}. Complementary information from micromagnetic simulations of hysteresis (discussed in Supplementary \autoref{suppSect:MumaxHysteresis} \cite{Supplement}) is indicated by background shadings corresponding to distinct spin textures.  With decreasing field (\autoref{fig:hysteresis}a and \ref{fig:hysteresis}c), the spin texture transitions from a quasi-single domain state (unshaded region) into the nucleation of magnetic vortex core-like topologies in the spin texture near the sample edge, around 14 kA/m (purple-shaded region). Further decreasing the field pushes the cores continuously away from edge until a sharp transition near 6.5 kA/m yields another sizeable jump in demagnetization and a single vortex state  in the cap of the permalloy mushroom and a quasi-uniform state in the stem akin to a thick domain wall (blue-shaded region). An additional transition occurs near 4 kA/m, where the vortex core jumps close to the center of the film, leaving a narrower domain wall in the stem (gray-shaded region). A final transition into a texture supporting three domains in the stem while maintaining a vortex configuration in the cap occurs infrequently during the data collection at very low bias field and can be identified from its distinct signature (details in Supplementary \autoref{suppSect:IndividualHysteresis} \cite{Supplement}).  When the field sweep direction is reversed at low bias field (\autoref{fig:hysteresis}b and \ref{fig:hysteresis}d), the vortex core trajectory reverses and it translates back towards the edge of the sample until it is annihilated (gray-shaded to unshaded field region). 

The frequency-dependence of the hysteresis loop shapes is notably different between the $I_z^{\mathrm{RF}}$- and $I_y^{\mathrm{RF}}$-driven torques.  First, we concentrate on the unshaded, higher bias field regions in \autoref{fig:hysteresis}.  On account of the strong easy plane shape anisotropy of the permalloy island, the $I_z^{\mathrm{RF}}$-driven signal (seen in \autoref{fig:hysteresis}a and \ref{fig:hysteresis}b) arises predominantly from conventional cross product torques; it is nearly frequency independent and approximately proportional to the net magnetization along the $x$-direction (as discussed in Supplementary \autoref{suppSect:MumaxHysteresis} \cite{Supplement}).  The unipolar hysteresis shape is commonly observed for cases of vortex nucleation and annihilation in confined micromagnets \cite{Cowburn1999}.  %[low bias field phase behavior of 64 MHz signal: most likely from the magnetic field dependence of the mechanical resonance frequency via the DC magnetic torque.  how could this be so much more visible just for this 64 MHz, $I_z$-driven data?  combination of mode Q and positioning of drive freq. relative to mech. res. freq. at high field?]  

The $I_y^{\mathrm{RF}}$-driven hysteresis, on the other hand, has a significant frequency dependence arising from EdH torques. Whereas the 3 MHz curve strongly resembles all $I_z^{\mathrm{RF}}$-driven curves (Supplementary \autoref{suppSect:3MHzHysteresis} \cite{Supplement}), higher frequency curves exhibit additional behaviours. The quasi-uniform magnetization regime persists over the widest field range in the hysteresis branch sweeping from high field to low field (\autoref{fig:hysteresis}a and \ref{fig:hysteresis}c), until nucleation of a vortex core-like object.  In the higher field region we observe a new contribution to the torque magnitude varying inversely with bias field and increasing with frequency, two of the features found in the macrospin simulations.  A surprise in these data is the near constancy (bias field independence) of the $I_y^{\mathrm{RF}}$-driven signal phases through the quasi-uniform magnetization state and across the spin texture transition at 14 kA/m in the field-decreasing hysteresis branch.  Similar to the macrospin simulation for mixed drive field components in \autoref{fig:macrospin}b, it was expected that bias field-dependent changes in the relative magnitudes of the EdH and cross product torques would yield a larger bias field-dependence of the phase of the net torque than we observe.  The predicted phase variation at 114 MHz is smaller than that shown for 208 MHz in \autoref{fig:macrospin}b, but still significantly larger than observed. %\kf{Add concluding sentence that mentions we continue this discussion in Section 3}

Below 6.5 kA/m in the decreasing field branch, spikes in torque related to the Barkhausen effect arise.  The high exchange energy density of the vortex core as well as its small diameter (approximately 10 nm) allows for high probability of interaction with pinning potentials created by magnetic disorder in the polycrystalline film.  If the pinning potentials are weak enough to be overcome by thermal fluctuations, the core can experience hopping between neighboring sites \cite{Burgess2013}.  The hopping rates at room temperature are generally high in comparison to the mechanical detection bandwidths.  The superposition of the random disorder potential for the core with continuous tilting from the external bias field can yield circumstances during the hysteresis sweep where the energy barrier is such that thermal activation assists the core in hopping synchronously with the in-plane drive. If the associated change in $y$-moment is larger than normal, this results in a spike in torque with a change in phase arising also through the thermal assist \cite{Hajisalem2019}.  These features recur across multiple hysteresis measurements when the same bias field conditions are reproduced.  Hysteresis in the low-field regime is isolated in Supplementary \autoref{suppFig:LowFieldHysteresis} to illustrate these features \cite{Supplement}.  

Whereas $H_y^{\mathrm{RF}}$ modulates the core position laterally along the sample surface, the $H_z^{\mathrm{RF}}$-drive minimally perturbs the core aside from small fluctuations of its width.  This is reflected in the absence of Barkhausen signatures in \autoref{fig:hysteresis}a and \ref{fig:hysteresis}b.  Some peaks seen in the 114 MHz signal are from the small admixture of cross product and EdH torques due to a non-zero component of in-plane ($H_y^{\mathrm{RF}}$) magnetic field in the $I_z^{\mathrm{RF}}$ drive (Supplementary \autoref{suppSect:RFAdmixture} \cite{Supplement}).  With increasing detection frequencies, the Barkhausen effect features become more pronounced and larger in magnitude with respect to the overall hysteresis curve, with corresponding increase in phase modulation.

\subsection{Measurements at 208 MHz mode: EdH torque at a gyrotropic resonance} \label{sect:208MHz_RFMeasurements}
\begin{figure}[ht] 
	\centering
    \includegraphics[width=1\linewidth] {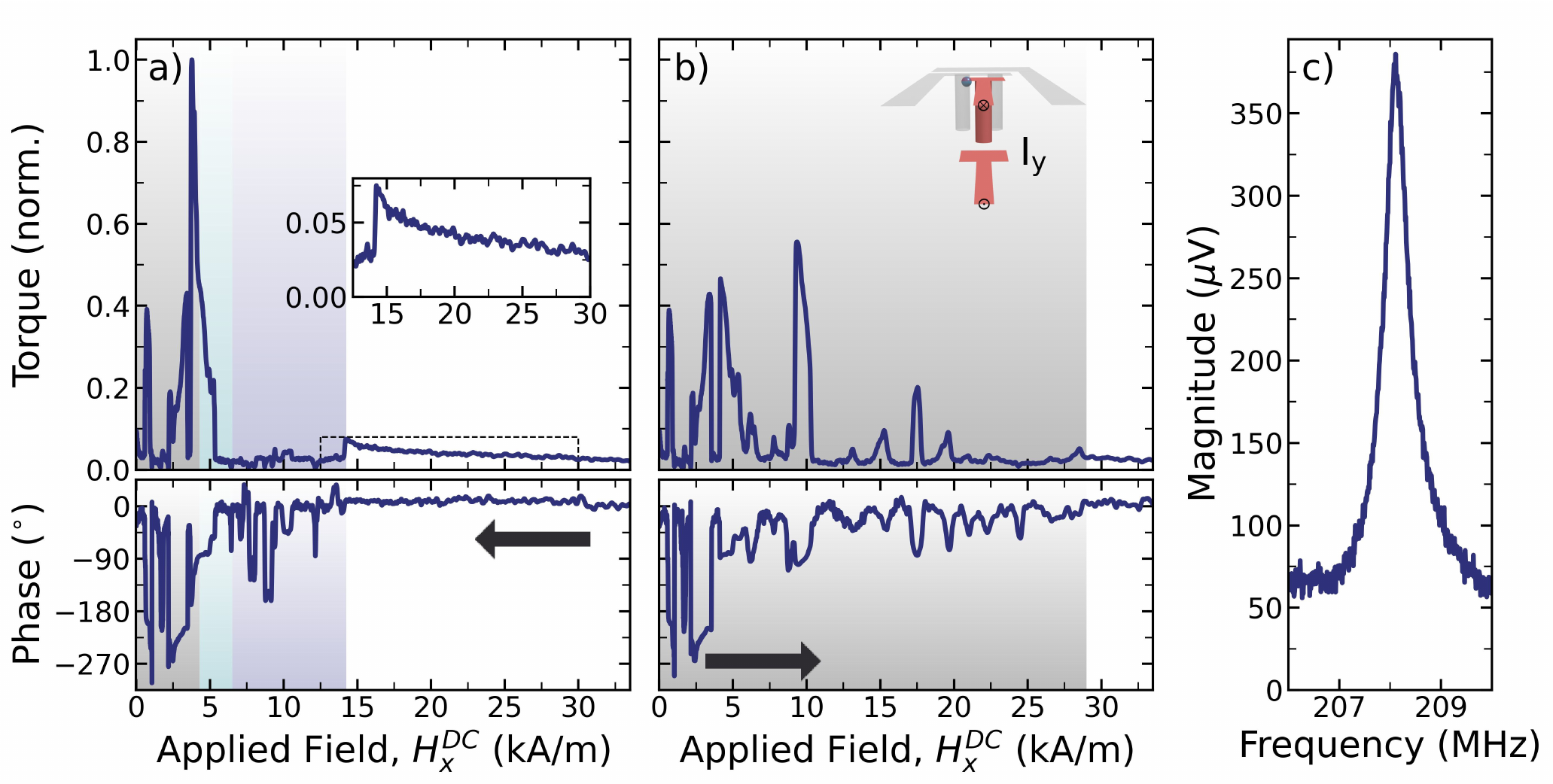}
	\caption{\textbf{Hysteresis of the 208 MHz mechanical mode through magnetic resonance.} $I_y^{\mathrm{RF}}$-driven hysteresis of the highest-frequency measured mode from (a) high to low (b) low to high field. The accompanying phase is shown below the magnitude. Distinct spin textures are indicated as in \autoref{fig:hysteresis}. The high-field regime ($H_x^{\mathrm{DC}}>$12.5 kA/m) is isolated in the inset of (a) to highlight the familiar hysteresis shape as seen in the lower-frequency $I_y^{\mathrm{RF}}$-driven hysteresis of \autoref{fig:hysteresis} (c) The mechanical resonance measured at the peak of the hysteresis loop (around 4 kA/m).}
	\label{fig:magnres}
% \mf{question: the data in c are most likely not background-subtracted, with the Lorentzian fit including an offset?  this will likely lead to confusion and consternation.  Do we have the phase information from that sweep through the mechanical resonance?  Is there a chance of performing a phasor background subtraction?}

  %indicated within the bandwidth of the resonance from simulation.
 
 %Micromagnetic simulation results of the single-vortex state at zero field are shown in (b).  The time-trace used for the FFT calculation (data points) is shown in the inset.  The mechanical resonance measured at the peak of the hysteresis (around 4 kA/m) is indicated within the bandwidth of the resonance from simulation. 
 \end{figure}

The $I_y^{\mathrm{RF}}$-driven magnetic hysteresis measured at the 208 MHz mechanical mode (\autoref{fig:mechmodes}e) in \autoref{fig:magnres} is at the highest frequency implemented for cantilever-style mechanical torque magnetometry to date.  Empirically, our torque detection sensitivity decreases approximately as $e^{-0.025f}$ with increasing frequency ($f$ in MHz).  The optimized signals for $I_z^{\mathrm{RF}}$ drives of 180 mV RMS from the lock-in amplifier through an ENI 510L power amplifier at $H_x =$ 53 kA/m were 1104 $\pm$ 8 $\mu V$, 656 $\pm$ 3 $\mu V$, 248 $\pm$ 1 $\mu V$, and 62.1 $\pm$ 0.9 $\mu V$ for the 3, 21, 64, and 114 MHz modes.  Further reduced mechanical transduction of the 208 MHz mode yielded difficulty in finding the signal through the usual procedure of optimizing the drive frequency, laser wavelength tuning, and fiber dimple positioning at high bias field with $I_z^{\mathrm{RF}}$ drive.  Nature did us a favour however and arranged for the presence of a huge torque signal at low bias field through an overlap of the 208 MHz mechanical mode with an in-plane RF field driven spin resonance in the permalloy island.  

The sweep from high to low field (\autoref{fig:magnres}a) follows a similar trend as seen for the in-plane $H_y^{\mathrm{RF}}$-driven modes in \autoref{fig:hysteresis}c until the spin texture nucleates a magnetic vortex.  In the vortex spin texture, the torque signal is substantially larger than that observed for the other measured in-plane $I_y^{\mathrm{RF}}$-driven modes, peaking at around 4 kA/m.  This is due to mechanical transduction of a spin resonance mode of the vortex texture, where the in-plane field drives the vortex core into a gyrotropic orbit \cite{Compton2006,Compton2010}. Abrupt reduction of the RF torque occurs at certain fields where the vortex core becomes strongly pinned by magnetic disorder in the sample.    For a vortex core in a thin film, out-of-plane fields do not contribute to translation of the vortex core and hence will not contribute to gyromagnetic precession. Therefore, no amplification of the 208 MHz $I_z^{\mathrm{RF}}$-driven torques is expected as the result of spin resonance. Paired with the poor mechanical transduction of the high-order resonant modes and high RF background, the cross product torque component of the $I_z^{\mathrm{RF}}$-driven torque is minimal (see Supplementary \autoref{suppSect:208MHzHzRF} \cite{Supplement}). 

\begin{figure}[ht]
    \centering
    \includegraphics[width=0.7\linewidth]{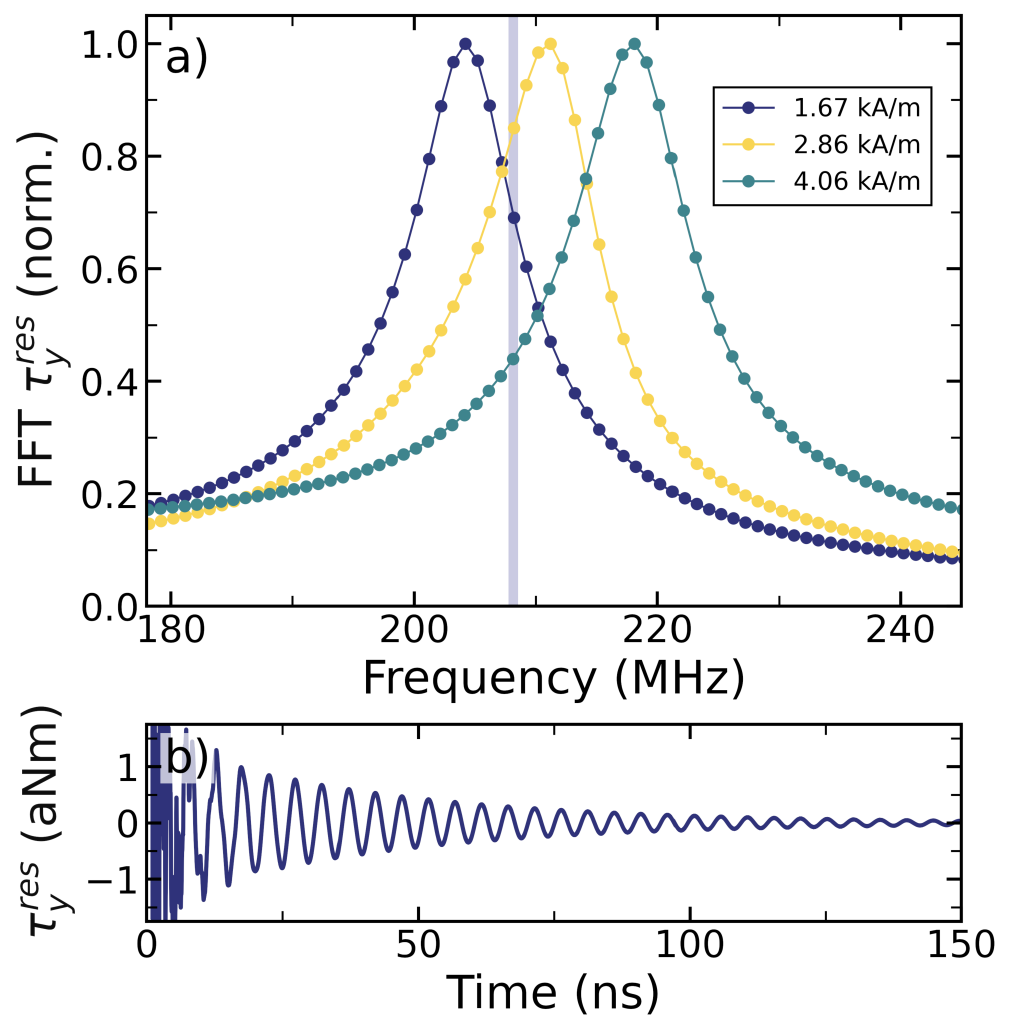}
    \caption{\textbf{Micromagnetic simulations of torque across the gyrotropic resonance, for several DC bias fields.} FFT calculations of micromagnetic simulation output for a 1 ns field pulse applied to the single-vortex state at three distinct bias fields are shown in (a). The shaded vertical bar indicates the frequency and linewidth of the measured mechanical resonance from \autoref{fig:magnres}b. The time trace of resultant torque at 1.67 kA/m bias field is shown in (b).}
    \label{fig:ringdowns}
\end{figure}

The existence of the gyrotropic mode was confirmed through a series of micromagnetic simulations of the sample geometry (performed with mumax$^3$, an open-source, GPU-based LLG solver \cite{Vansteenkiste2014}).  The spin texture at 1.67 kA/m was first acquired by running a hysteresis simulation from high to low field to promote spin textures seen in experiment.  In a separate simulation, a 1 ns wide, 0.398 A/m Heaviside square pulse in the plane of the sample was applied to the spin texture and the magnetization was allowed to relax.  The FFT magnitude (\autoref{fig:ringdowns}a) of the resulting time trace of the net $y$-torque (conventional plus EdH torques, \autoref{fig:ringdowns}b) demonstrates a broad magnetic resonance which overlaps the narrow bandwidth of measured torque (shaded band in \autoref{fig:ringdowns}a). The coincidence of magnetic and mechanical resonances persists over a wide applied field range, as indicated by repeating the aforementioned simulations at 2.86 kA/m and 4.06 kA/m (\autoref{fig:ringdowns}a). This field range encompasses the experimental field range wherein a single vortex state is roughly centered in the permalloy mushroom. %Experimental indication of the gyromagnetic mode is achieved through torque-mixing resonant spectroscopy \cite{Losby2015} (details in Supplementary \autoref{suppSect:TMRS}).

Gyrotropic frequencies in polycrystalline permalloy microstructures are strongly modified by the imprint of magnetic disorder on the energy of the vortex core versus position, aided by field-tuning of the resonance (a gradual change of the gyrotropic frequency with field in the vortex state, and enhanced by interaction of the vortex core with pinning sites in the permalloy).  Local pinning potentials are caused by variations in film thickness due to granularity and in the strength of exchange coupling between grains.  The frequencies of small amplitude gyrotropic orbits increase on account of the local curvature of core energy versus position, which can be significantly larger than the global curvature owing to the overall shape of the magnetic structure and of the spin configuration within it.  Mapping the gyrotropic frequency versus applied field then yields a map of the static, random potential.  If the pinning is not too strong then at some bias fields, and for strong enough RF drive, the core will manage to surf over the disorder at a gyrotropic frequency close to that expected for a perfectly uniform film. 

% Experimentally, RF drive thresholds for core depinning and high gyrotropic frequencies within pinning sites were first studied in time-resolved magneto-optical imaging measurements performed by the Crowell group \cite{Compton2010}.  In the present work we also observe drive thresholds for the strong torque signals, and have detected the gyrotropic mode independently through torque-mixing magnetic resonance spectroscopy (details in Supplementary \autoref{suppSect:DriveThreshold} and \autoref{suppSect:TMRS}, respectively). \kf{Revisit this comment re. ski slope onset in Fig S10 }

\begin{figure}[ht]
    \centering
    \includegraphics[width=0.95\linewidth]{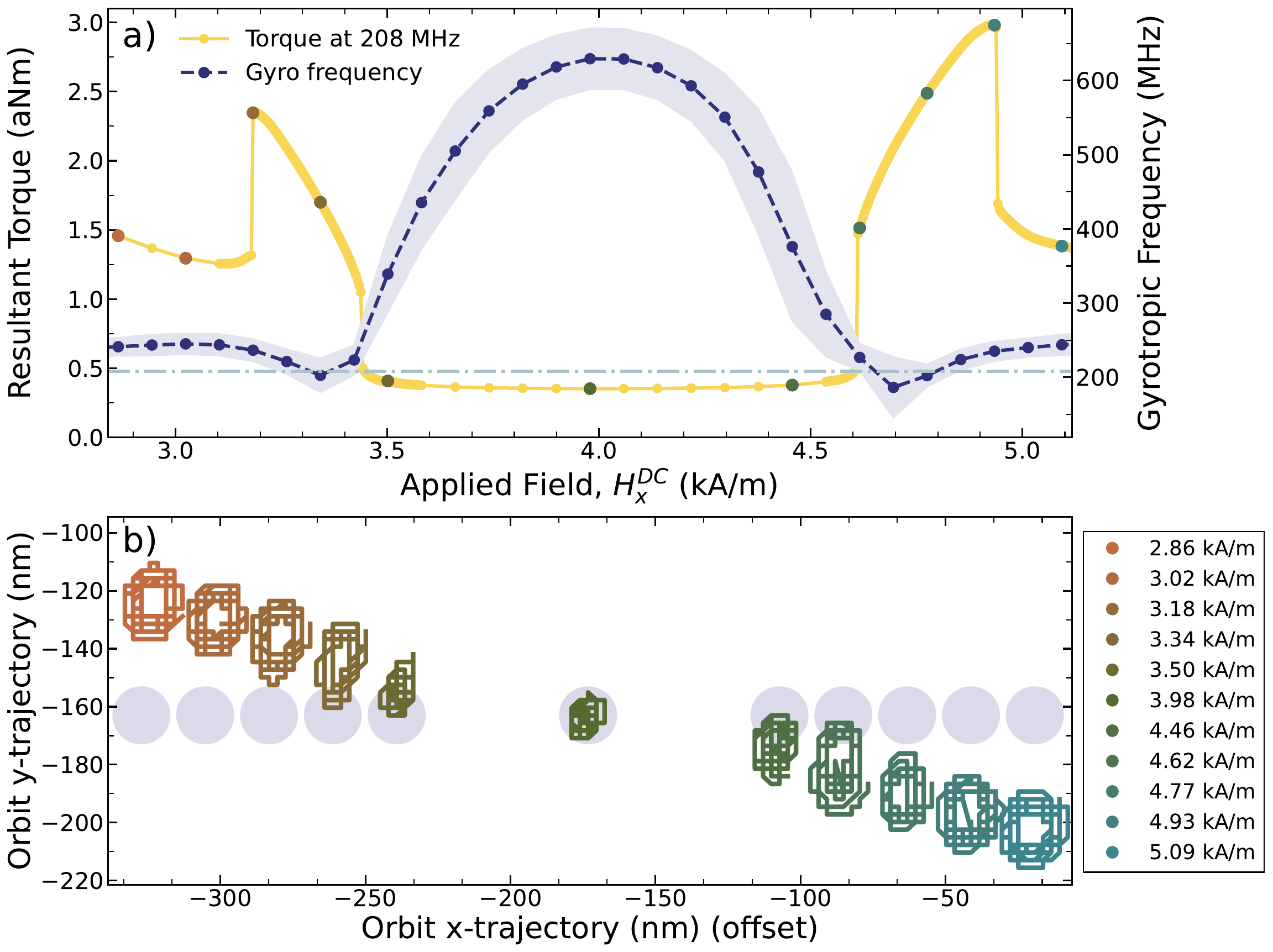}
    \caption{\textbf{Simulated $\mathrm{H_y^{RF}}$-driven torques at 208 MHz in the vicinity of a vortex core pinning site.} (a) Simulated torque (yellow curve) from low to high field for a time evolving micromagnetic simulation with a defect positioned in the path of the vortex core. The local gyrotropic frequency, shown in purple and plotted against the right-hand vertical axis, is strongly and non-monotonically position (bias field) dependent outside of and within the pinning site.  The 208 MHz driving frequency is indicated by the horizontal blue line.  As an aid in visualizing the origin of the resultant, predominantly EdH torque, the core trajectories for eleven of the specific driven gyrotropic motions are shown in (b), corresponding to the discrete field points indicated on the torque curve in (a). The defect location is offset along the $x$-axis for each DC field value, indicated by light blue circles. }
    \label{fig:gyrohysteresis}
\end{figure}

We further elucidate the specific effects of pinning and depinning of the gyrotropic mode on magnetic torque through simulations from low to high field with a 20 nm diameter cylindrical defect positioned along the trajectory of the vortex core within the permalloy island (see Supplementary \autoref{suppSect:PinnedSimulations} for simulation details \cite{Supplement}). The defect mimics magnetic disorder using a slightly reduced $M_s$ for the defect, providing a lower energy position for the vortex core to become trapped at over a small bias field range. As the core approaches the pinning site, the gyrotropic orbit elongates and the frequency decreases, as seen in \autoref{fig:gyrohysteresis}. The energy minimum of the pinning site behaves like an attractor and adds additional angular momentum to the core as it nears the defect, in an approximate parallel to a gravitational slingshot. Direct interaction with the pinning site reduces angular momentum (and therefore the torque) due to the energy cost required to exit the defect's influence. The reduced energy requirements inside the defect leads to a significant increase in gyrotropic frequency, with maximum frequency occurring when the orbit is centred within the defect. The effect of an additional pinning site along the core's trajectory is explored in Supplementary \autoref{suppSect:DoublePinnedSimulations} wherein general agreement of phase and magnitude behaviour between measured and simulated torques is observed \cite{Supplement}. 

\section{Global model of torque magnitudes and phases in the quasi-uniform spin texture}

The admixture of $z$-field from the $\mathrm{H_y^{RF}}$ coil has a prominent effect on measured torques.  COMSOL simulation of the field produced by the coil geometry (Fig. 3) indicates an opposite sign of $\mathrm{H_z^{RF}}$ relative to $\mathrm{H_y^{RF}}$ at the sample position when driving the $\mathrm{H_y^{RF}}$ coil.  Because of the importance of signal phases, this must be incorporated into micromagnetic simulations by including the sign of the field ratio and computing the superposition of $ -H_z$-driven torques, scaled by the field ratio, with $\mathrm{H_y}$-driven torques. The effect of this is evident in Fig 8, in which the high-field portion of the measured hysteresis is isolated and compared to these simulations. The simulated, admixed signal phase slopes downwards to higher bias fields, in qualitative agreement with the phase slopes observed in the measured torques (opposite to what was seen for a positive ratio admixture in \autoref{fig:macrospin}).  

\begin{figure}[ht] 
	\centering
    \includegraphics[width=0.7\linewidth]{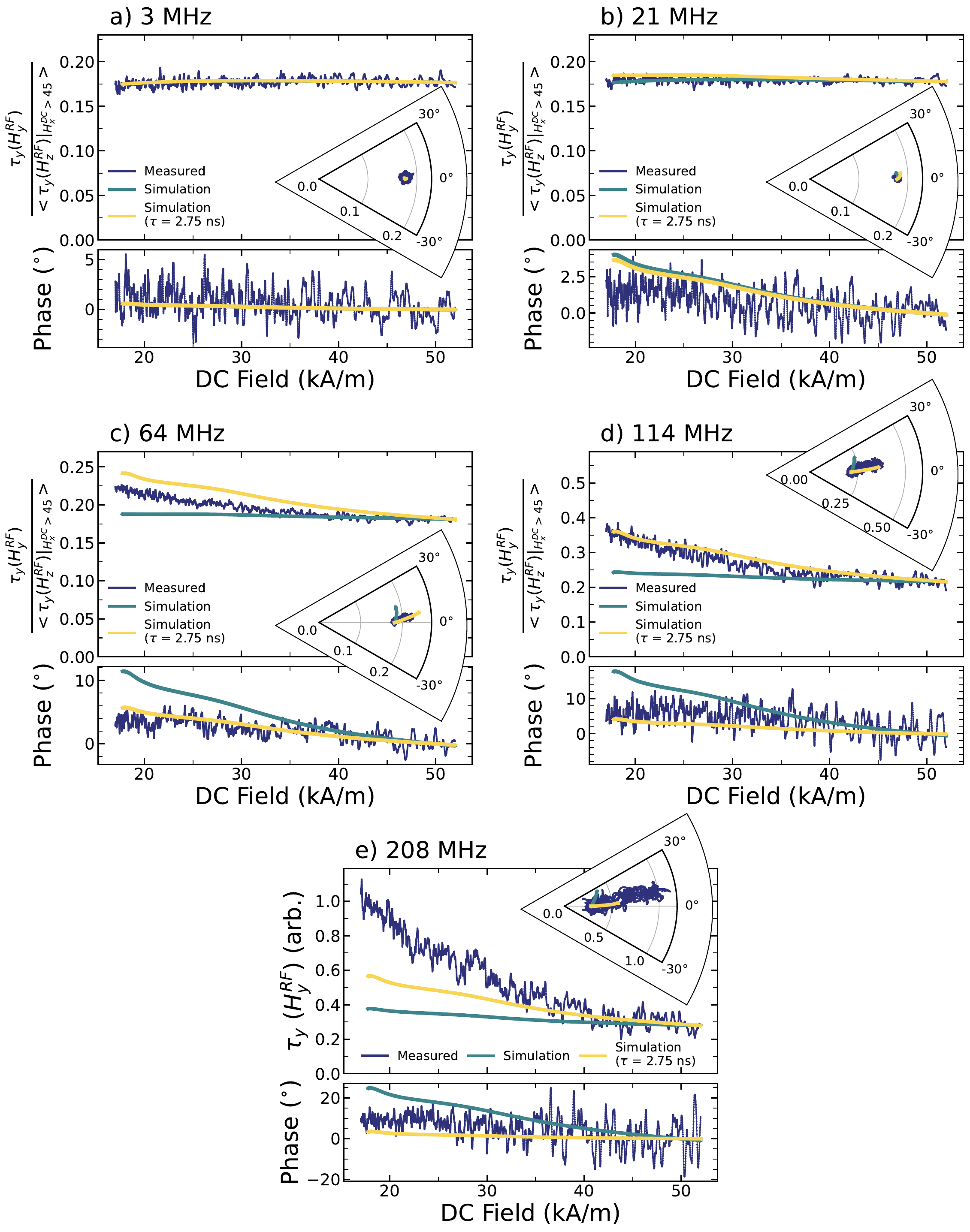}
    \caption{\textbf{High-field hysteresis and micromagnetic simulation.} Experimental torque signals (purple) as presented in Figure 3 in the main text compared against simulation output for primarily $H_y^{\mathrm{RF}}$-driven torques superimposed with a small -$H_z^{\mathrm{RF}}$ drive contribution (-25.2\% of the $H_y^{\mathrm{RF}}$ contribution). The opposite sign of $H_z^{\mathrm{RF}}$ and $H_y^{\mathrm{RF}}$ is indicated by simulation of the field vector at the sample location (\autoref{fig:fieldGeometry}), contrary to macrospin simulation output presented in \autoref{fig:macrospin}b where the admixed field components have the same relative phase. An exponentially weighted mean with time constant $\tau$ = 2.75 ns is applied to the $H_z^{\mathrm{RF}}$-driven simulation torque (yellow) to mimic the effect of a relaxation process in the cross product torque. This relaxation process empirically improves agreement to measured torques in comparison to the non-filtered combination (green). Insets of polar segments provide a alternate view of the experiment/model comparison, illustrating the heightened agreement through filtering.}
    \label{fig:SkiSlopeAnalysis}
\end{figure}

A global comparison at all frequencies of the raw, simulated admixture torques shows poor quantitative correspondence to measured torques over the 14 to 53 kA/m field range, corresponding to the quasi-uniform spin texture field-decreasing branch (\autoref{fig:SkiSlopeAnalysis}).  At the upper field limit of this spin texture, the $\mathrm{H_z}$-driven torques dominate and admixture of $\mathrm{H_y}$ contributes little to the torque. Closer to nucleation of a vortex core at the lower field limit of the quasi-uniform spin texture, the $\mathrm{H_y}$-driven torques play a larger role. In this region the correspondence between simulation and experiment weakens.  The quantitative distinctions, which show up with greater contrast at higher frequencies where the EdH torque contribution is larger, are that the experimental phase has less overall field slope than simulation, and that the simulated, resultant torque magnitudes have less overall field slope than found from experiment.  Taken together, these quantitative differences point in the direction of a breakdown of the strict phase orthogonality ($\pi/2$ relative phase difference) between the  EdH and cross product torques within the frequency range of this study.  In the phasor superposition of a smaller but growing torque component with a larger and constant component, the overall magnitude (phase) variation will increase (decrease) when the relative phase difference between the two moves away from $\pi/2$. A likely cause for this behaviour is the decoupling of internal phonon modes from the rigid body rotation of the resonator \cite{Ruckriegel2020}. 

Timescales for magnetic torques are a topic of considerable current interest.  With ultrafast x-ray observations of phonon generation by EdH torque on sub-picosecond timescales \cite{Dornes2019}, we assume that the EdH mechanical torques are effectively instantaneous at our frequencies of measurement (although the question of bottlenecks due to the cascade from short- to long-wavelength phonons remains open).  The timescale for mechanical torque generation from shape anisotropy alone (the composition of Ni$_{80}$Fe$_{20}$ is tailored to minimize magnetoelastic response) are undetermined, to our knowledge.  Phenomenologically, a single pole low-pass filter is the simplest model addition to add phase-shifting between a cross product magnetic shape anisotropy torque and its resulting rigid body mechanical torque.  In an circuit analogy, we envision the RF magnetic torque as an AC current source driving a parallel $RC$ circuit, where the  $RC$ integrator time constant mimics the time-dependence of torque conversion and the current in the resistor represents the rigid body mechanical torque.  

%\mf{have we mentioned somewhere -- sniffer coil section of supplement? -- estimate of eddy current decay time in the permalloy for $H_z$ drive}

The $RC$ circuit model is included using a circuit time constant of 2.75 ns (within the range of 0.5 - 4 ns constrained by experiment). This model is applied to simulations via an exponentially weighted mean function acting on the $H_z$-driven torque to act as the $RC$ filter. The filtered $H_z$-driven torques are superimposed with a $H_y$-driven torque with relative drive field strengths in agreement with COMSOL simulations of the driving coils. This procedure pushes simulations to closer agreement with experiment (\autoref{fig:SkiSlopeAnalysis}) by increasing (decreasing) the overall slope of the magnitude (phase). This agreement is observed plainly in the polar insets of \autoref{fig:SkiSlopeAnalysis} which show the filtered simulations mapping directly to measurement, contrary to behaviour observed in unfiltered simulations.  The 208 MHz simulations have notably worse agreement in torque magnitude than the lower order modes. This can be attributed in part to softening of the spin texture before nucleation of the vortex, as closure domains near the edges of the mushroom shape become trapped. This trapping produces different resonance behaviour than expected for a uniformly magnetized sample. Simulation of FMR of the mushroom shape in the 18 to 25 kA/m field range supports this explanation, revealing broad frequency signatures near 200 MHz for an excitation in the $H_y$ drive which will interact with the 208 MHz drive frequency.

%\textbf{ Discussion:}
%a likely feature of the admixture of HxRF from the y-coil is that it should guarantee a DC field-independent phase between the two field drive components at the sample.  (likely even in phase with each other, although that is not guaranteed because of possible screening currents.  but there is no apparent mechanism for the phase relationship to change with DC field.)  Future experiments also need to be performed at low-enough temperatures to simplify the observations by eliminating thermally-activated, slower dynamics.  The ability to change the sample position relative to the excitation coils will be very important in ongoing studies.  For example, recording the torque magnitude and phase after translating the sample across the y-coil to change the sign of the $\mathrm{H_z^{RF}}$ admixture (equivalent to changing its phase by 180$^{\circ}$) will provide an independent measure of the relative phase between the $\mathrm{H_y}$- and $\mathrm{H_z}$-driven torque responses.  

\section{Conclusion}

To summarize, nanocavity optomechanical sensing extends the mechanical detection (and corresponding AC magnetic drive) frequency range beyond previous studies by almost two orders of magnitude, in the measurements reported here.  This opens two additional regimes to EdH observations: torques generated by AC perturbation of the direction of the moment in magnetically-saturated specimens; and nonequilibrium torques boosted by a collective mode of the magnetization at the mechanical detection frequency.  

Current experimental limitations prohibit sample translation relative to the excitation coils. Ongoing studies will incorporate such adjustments to provide independent measure of the relative phase between the $H_y$- and $H_z$-driven torque responses, e.g. by recording the torque magnitude and phase after translating the sample across the y-coil to change the sign of the $H_z^{\mathrm{RF}}$ admixture (equivalent to changing its phase by 180$^{\circ}$). Determination of the absolute phases of the drive fields at the sample location will also be valuable, and can be accomplished using a magneto-optical probe. Employing a phase-locked loop during measurement will prevent any thermally induced frequency shifts from insidiously affecting measured torques. Including an additional axis of driving field would allow further excitation of the gyrotropic mode to expand this study with the caveat that, with the current field geometry, no further low-frequency torques could be excited in this manner. Adjusting the field geometry or utilizing a sensor boasting mechanical compliance to an additional torque axis are reasonable solutions to be employed in related studies.  Additionally, future work will be performed at low temperatures to eliminate the slower, thermally activated dynamics and thereby simplifying the observations.

\begin{acknowledgements}
The authors gratefully acknowledge support from the Natural Sciences and Engineering Research Council, Canada (RGPIN 2021-02762), the Canada Foundation for Innovation (34028), the Canada Research Chairs (230377), the National Research Council (Canada), and the University of Alberta.  
\end{acknowledgements}

\textbf{Code Availability:} The code that supports this work including the macrospin simulation, lock-in emulator, MuMax3 scripts, and processing scripts available via GitHub at \url{https://github.com/KatrynaFast/EdH_DiscreteSpectroscopicProbe_Analysis-info}. 

%\bibliography{refs.bib}

\begin{thebibliography}{37}%
\makeatletter
\providecommand \@ifxundefined [1]{%
 \@ifx{#1\undefined}
}%
\providecommand \@ifnum [1]{%
 \ifnum #1\expandafter \@firstoftwo
 \else \expandafter \@secondoftwo
 \fi
}%
\providecommand \@ifx [1]{%
 \ifx #1\expandafter \@firstoftwo
 \else \expandafter \@secondoftwo
 \fi
}%
\providecommand \natexlab [1]{#1}%
\providecommand \enquote  [1]{``#1''}%
\providecommand \bibnamefont  [1]{#1}%
\providecommand \bibfnamefont [1]{#1}%
\providecommand \citenamefont [1]{#1}%
\providecommand \href@noop [0]{\@secondoftwo}%
\providecommand \href [0]{\begingroup \@sanitize@url \@href}%
\providecommand \@href[1]{\@@startlink{#1}\@@href}%
\providecommand \@@href[1]{\endgroup#1\@@endlink}%
\providecommand \@sanitize@url [0]{\catcode `\\12\catcode `\$12\catcode
  `\&12\catcode `\#12\catcode `\^12\catcode `\_12\catcode `\%12\relax}%
\providecommand \@@startlink[1]{}%
\providecommand \@@endlink[0]{}%
\providecommand \url  [0]{\begingroup\@sanitize@url \@url }%
\providecommand \@url [1]{\endgroup\@href {#1}{\urlprefix }}%
\providecommand \urlprefix  [0]{URL }%
\providecommand \Eprint [0]{\href }%
\providecommand \doibase [0]{http://dx.doi.org/}%
\providecommand \selectlanguage [0]{\@gobble}%
\providecommand \bibinfo  [0]{\@secondoftwo}%
\providecommand \bibfield  [0]{\@secondoftwo}%
\providecommand \translation [1]{[#1]}%
\providecommand \BibitemOpen [0]{}%
\providecommand \bibitemStop [0]{}%
\providecommand \bibitemNoStop [0]{.\EOS\space}%
\providecommand \EOS [0]{\spacefactor3000\relax}%
\providecommand \BibitemShut  [1]{\csname bibitem#1\endcsname}%
\let\auto@bib@innerbib\@empty
%</preamble>
\bibitem [{\citenamefont {Einstein}\ and\ \citenamefont
  {de~Haas}(1915)}]{Einstein1915}%
  \BibitemOpen
  \bibfield  {author} {\bibinfo {author} {\bibfnamefont {A.}~\bibnamefont
  {Einstein}}\ and\ \bibinfo {author} {\bibfnamefont {W.~J.}\ \bibnamefont
  {de~Haas}},\ }in\ \href@noop {} {\emph {\bibinfo {booktitle} {KNAW,
  Proceedings}}},\ Vol.\ \bibinfo {volume} {18 I}\ (\bibinfo {year} {1915})\
  p.\ \bibinfo {pages} {696}\BibitemShut {NoStop}%
\bibitem [{\citenamefont {Wallis}\ \emph {et~al.}(2006)\citenamefont {Wallis},
  \citenamefont {Moreland},\ and\ \citenamefont {Kabos}}]{Wallis2006}%
  \BibitemOpen
  \bibfield  {author} {\bibinfo {author} {\bibfnamefont {T.~M.}\ \bibnamefont
  {Wallis}}, \bibinfo {author} {\bibfnamefont {J.}~\bibnamefont {Moreland}}, \
  and\ \bibinfo {author} {\bibfnamefont {P.}~\bibnamefont {Kabos}},\ }\href
  {https://doi.org/10.1063/1.2355445} {\bibfield  {journal} {\bibinfo
  {journal} {Appl. Phys. Lett.}\ }\textbf {\bibinfo {volume} {89}},\ \bibinfo
  {pages} {122502} (\bibinfo {year} {2006})}\BibitemShut {NoStop}%
\bibitem [{\citenamefont {Jaafar}\ \emph {et~al.}(2009)\citenamefont {Jaafar},
  \citenamefont {Chudnovsky},\ and\ \citenamefont {Garanin}}]{Jaafar2009}%
  \BibitemOpen
  \bibfield  {author} {\bibinfo {author} {\bibfnamefont {R.}~\bibnamefont
  {Jaafar}}, \bibinfo {author} {\bibfnamefont {E.~M.}\ \bibnamefont
  {Chudnovsky}}, \ and\ \bibinfo {author} {\bibfnamefont {D.~A.}\ \bibnamefont
  {Garanin}},\ }\href {\doibase .1103/PhysRevB.79.10441100} {\bibfield
  {journal} {\bibinfo  {journal} {Phys. Rev. B}\ }\textbf {\bibinfo {volume}
  {79}},\ \bibinfo {pages} {104410} (\bibinfo {year} {2009})}\BibitemShut
  {NoStop}%
\bibitem [{\citenamefont {Chudnovsky}\ and\ \citenamefont
  {Garanin}(2014)}]{Chudnovsky2014}%
  \BibitemOpen
  \bibfield  {author} {\bibinfo {author} {\bibfnamefont {E.~M.}\ \bibnamefont
  {Chudnovsky}}\ and\ \bibinfo {author} {\bibfnamefont {D.~A.}\ \bibnamefont
  {Garanin}},\ }\href {\doibase 10.1103/PhysRevB.89.174420} {\bibfield
  {journal} {\bibinfo  {journal} {Phys. Rev. B}\ }\textbf {\bibinfo {volume}
  {89}},\ \bibinfo {pages} {174420} (\bibinfo {year} {2014})}\BibitemShut
  {NoStop}%
\bibitem [{\citenamefont {Zarzuela}\ and\ \citenamefont
  {Chudnovsky}(2015)}]{Zarzuela2015}%
  \BibitemOpen
  \bibfield  {author} {\bibinfo {author} {\bibfnamefont {R.}~\bibnamefont
  {Zarzuela}}\ and\ \bibinfo {author} {\bibfnamefont {E.~M.}\ \bibnamefont
  {Chudnovsky}},\ }\href {\doibase 10.1007/s10948-015-3184-5} {\bibfield
  {journal} {\bibinfo  {journal} {Journal of Superconductivity and Novel
  Magnetism}\ }\textbf {\bibinfo {volume} {28}},\ \bibinfo {pages} {3411}
  (\bibinfo {year} {2015})}\BibitemShut {NoStop}%
\bibitem [{\citenamefont {Wells}\ \emph {et~al.}(2019)\citenamefont {Wells},
  \citenamefont {Horsfield}, \citenamefont {Foulkes},\ and\ \citenamefont
  {Dudarev}}]{Wells2019}%
  \BibitemOpen
  \bibfield  {author} {\bibinfo {author} {\bibfnamefont {T.}~\bibnamefont
  {Wells}}, \bibinfo {author} {\bibfnamefont {A.~P.}\ \bibnamefont
  {Horsfield}}, \bibinfo {author} {\bibfnamefont {W.~M.}\ \bibnamefont
  {Foulkes}}, \ and\ \bibinfo {author} {\bibfnamefont {S.~L.}\ \bibnamefont
  {Dudarev}},\ }\href {\doibase 10.1063/1.5092223} {\bibfield  {journal}
  {\bibinfo  {journal} {J. Chem. Phys.}\ }\textbf {\bibinfo {volume} {150}},\
  \bibinfo {pages} {224109} (\bibinfo {year} {2019})}\BibitemShut {NoStop}%
\bibitem [{\citenamefont {R{\"{u}}ckriegel}\ \emph {et~al.}(2020)\citenamefont
  {R{\"{u}}ckriegel}, \citenamefont {Streib}, \citenamefont {Bauer},\ and\
  \citenamefont {Duine}}]{Ruckriegel2020}%
  \BibitemOpen
  \bibfield  {author} {\bibinfo {author} {\bibfnamefont {A.}~\bibnamefont
  {R{\"{u}}ckriegel}}, \bibinfo {author} {\bibfnamefont {S.}~\bibnamefont
  {Streib}}, \bibinfo {author} {\bibfnamefont {G.~E.~W.}\ \bibnamefont
  {Bauer}}, \ and\ \bibinfo {author} {\bibfnamefont {R.~A.}\ \bibnamefont
  {Duine}},\ }\href {\doibase 10.1103/PhysRevB.101.104402} {\bibfield
  {journal} {\bibinfo  {journal} {Phys. Rev. B}\ }\textbf {\bibinfo {volume}
  {101}},\ \bibinfo {pages} {104402} (\bibinfo {year} {2020})}\BibitemShut
  {NoStop}%
\bibitem [{\citenamefont {Mori}\ \emph {et~al.}(2020)\citenamefont {Mori},
  \citenamefont {Dunsmore}, \citenamefont {Losby}, \citenamefont {Jenson},
  \citenamefont {Belov},\ and\ \citenamefont {Freeman}}]{Mori2020}%
  \BibitemOpen
  \bibfield  {author} {\bibinfo {author} {\bibfnamefont {K.}~\bibnamefont
  {Mori}}, \bibinfo {author} {\bibfnamefont {M.~G.}\ \bibnamefont {Dunsmore}},
  \bibinfo {author} {\bibfnamefont {J.~E.}\ \bibnamefont {Losby}}, \bibinfo
  {author} {\bibfnamefont {D.~M.}\ \bibnamefont {Jenson}}, \bibinfo {author}
  {\bibfnamefont {M.}~\bibnamefont {Belov}}, \ and\ \bibinfo {author}
  {\bibfnamefont {M.~R.}\ \bibnamefont {Freeman}},\ }\href
  {https://doi.org/10.1103/PhysRevB.102.054415} {\bibfield  {journal} {\bibinfo
   {journal} {Phys. Rev. B}\ }\textbf {\bibinfo {volume} {102}},\ \bibinfo
  {pages} {054415} (\bibinfo {year} {2020})}\BibitemShut {NoStop}%
\bibitem [{\citenamefont {Garanin}\ and\ \citenamefont
  {Chudnovsky}(2021)}]{Garanin2021}%
  \BibitemOpen
  \bibfield  {author} {\bibinfo {author} {\bibfnamefont {D.~A.}\ \bibnamefont
  {Garanin}}\ and\ \bibinfo {author} {\bibfnamefont {E.~M.}\ \bibnamefont
  {Chudnovsky}},\ }\href@noop {} {\bibfield  {journal} {\bibinfo  {journal}
  {Phys. Rev. B}\ }\textbf {\bibinfo {volume} {103}},\ \bibinfo {pages}
  {L100412} (\bibinfo {year} {2021})}\BibitemShut {NoStop}%
\bibitem [{\citenamefont {Dednam}\ \emph {et~al.}(2022)\citenamefont {Dednam},
  \citenamefont {Sabater}, \citenamefont {Botha}, \citenamefont {Lombardi},
  \citenamefont {Fern{\'{a}}ndez-Rossier},\ and\ \citenamefont
  {Caturla}}]{Dednam2022}%
  \BibitemOpen
  \bibfield  {author} {\bibinfo {author} {\bibfnamefont {W.}~\bibnamefont
  {Dednam}}, \bibinfo {author} {\bibfnamefont {C.}~\bibnamefont {Sabater}},
  \bibinfo {author} {\bibfnamefont {A.~E.}\ \bibnamefont {Botha}}, \bibinfo
  {author} {\bibfnamefont {E.~B.}\ \bibnamefont {Lombardi}}, \bibinfo {author}
  {\bibfnamefont {J.}~\bibnamefont {Fern{\'{a}}ndez-Rossier}}, \ and\ \bibinfo
  {author} {\bibfnamefont {M.~J.}\ \bibnamefont {Caturla}},\ }\href {\doibase
  10.1016/j.commatsci.2022.111359} {\bibfield  {journal} {\bibinfo  {journal}
  {Computational Materials Science}\ }\textbf {\bibinfo {volume} {209}},\
  \bibinfo {pages} {111359} (\bibinfo {year} {2022})}\BibitemShut {NoStop}%
\bibitem [{\citenamefont {Brooks}\ \emph {et~al.}(1987)\citenamefont {Brooks},
  \citenamefont {Naughton}, \citenamefont {Ma}, \citenamefont {Chaikin},\ and\
  \citenamefont {Chamberlin}}]{Brooks1987}%
  \BibitemOpen
  \bibfield  {author} {\bibinfo {author} {\bibfnamefont {J.~S.}\ \bibnamefont
  {Brooks}}, \bibinfo {author} {\bibfnamefont {M.~J.}\ \bibnamefont
  {Naughton}}, \bibinfo {author} {\bibfnamefont {Y.~P.}\ \bibnamefont {Ma}},
  \bibinfo {author} {\bibfnamefont {P.~M.}\ \bibnamefont {Chaikin}}, \ and\
  \bibinfo {author} {\bibfnamefont {R.~V.}\ \bibnamefont {Chamberlin}},\ }\href
  {https://doi.org/10.1063/1.1139552} {\bibfield  {journal} {\bibinfo
  {journal} {Rev Sci Instrum}\ }\textbf {\bibinfo {volume} {58}},\ \bibinfo
  {pages} {117} (\bibinfo {year} {1987})}\BibitemShut {NoStop}%
\bibitem [{\citenamefont {Cleland}\ and\ \citenamefont
  {Roukes}(1996)}]{Cleland1996}%
  \BibitemOpen
  \bibfield  {author} {\bibinfo {author} {\bibfnamefont {A.~N.}\ \bibnamefont
  {Cleland}}\ and\ \bibinfo {author} {\bibfnamefont {M.~L.}\ \bibnamefont
  {Roukes}},\ }\href {\doibase 10.1063/1.117548} {\bibfield  {journal}
  {\bibinfo  {journal} {Appl. Phys. Lett.}\ }\textbf {\bibinfo {volume} {69}},\
  \bibinfo {pages} {2653} (\bibinfo {year} {1996})}\BibitemShut {NoStop}%
\bibitem [{\citenamefont {Carr}\ and\ \citenamefont
  {Craighead}(1997)}]{Carr1997}%
  \BibitemOpen
  \bibfield  {author} {\bibinfo {author} {\bibfnamefont {D.~W.}\ \bibnamefont
  {Carr}}\ and\ \bibinfo {author} {\bibfnamefont {H.}~\bibnamefont
  {Craighead}},\ }\href {\doibase 10.1116/1.589722} {\bibfield  {journal}
  {\bibinfo  {journal} {J. Vac. Sci. Technol. B}\ }\textbf {\bibinfo {volume}
  {15}},\ \bibinfo {pages} {2760} (\bibinfo {year} {1997})}\BibitemShut
  {NoStop}%
\bibitem [{\citenamefont {Alzetta}\ \emph {et~al.}(1999)\citenamefont
  {Alzetta}, \citenamefont {Ascoli}, \citenamefont {Baschieri}, \citenamefont
  {Bertolini}, \citenamefont {Betti}, \citenamefont {Masi}, \citenamefont
  {Frediani}, \citenamefont {Lenci}, \citenamefont {Martinelli},\ and\
  \citenamefont {Scalari}}]{Alzetta1999}%
  \BibitemOpen
  \bibfield  {author} {\bibinfo {author} {\bibfnamefont {G.}~\bibnamefont
  {Alzetta}}, \bibinfo {author} {\bibfnamefont {C.}~\bibnamefont {Ascoli}},
  \bibinfo {author} {\bibfnamefont {P.}~\bibnamefont {Baschieri}}, \bibinfo
  {author} {\bibfnamefont {D.}~\bibnamefont {Bertolini}}, \bibinfo {author}
  {\bibfnamefont {I.}~\bibnamefont {Betti}}, \bibinfo {author} {\bibfnamefont
  {B.~D.}\ \bibnamefont {Masi}}, \bibinfo {author} {\bibfnamefont
  {C.}~\bibnamefont {Frediani}}, \bibinfo {author} {\bibfnamefont
  {L.}~\bibnamefont {Lenci}}, \bibinfo {author} {\bibfnamefont
  {M.}~\bibnamefont {Martinelli}}, \ and\ \bibinfo {author} {\bibfnamefont
  {G.}~\bibnamefont {Scalari}},\ }\href {\doibase 10.1006/jmre.1999.1878}
  {\bibfield  {journal} {\bibinfo  {journal} {Journal of Magnetic Resonance}\
  }\textbf {\bibinfo {volume} {141}},\ \bibinfo {pages} {148} (\bibinfo {year}
  {1999})}\BibitemShut {NoStop}%
\bibitem [{\citenamefont {Moreland}\ \emph {et~al.}(2001)\citenamefont
  {Moreland}, \citenamefont {Jander}, \citenamefont {Beall}, \citenamefont
  {Kabos},\ and\ \citenamefont {Russek}}]{Moreland2001}%
  \BibitemOpen
  \bibfield  {author} {\bibinfo {author} {\bibfnamefont {J.}~\bibnamefont
  {Moreland}}, \bibinfo {author} {\bibfnamefont {A.}~\bibnamefont {Jander}},
  \bibinfo {author} {\bibfnamefont {J.~A.}\ \bibnamefont {Beall}}, \bibinfo
  {author} {\bibfnamefont {P.}~\bibnamefont {Kabos}}, \ and\ \bibinfo {author}
  {\bibfnamefont {S.~E.}\ \bibnamefont {Russek}},\ }\href {\doibase
  10.1109/20.951302} {\bibfield  {journal} {\bibinfo  {journal} {IEEE
  Transactions on Magnetics}\ }\textbf {\bibinfo {volume} {37}},\ \bibinfo
  {pages} {2770} (\bibinfo {year} {2001})}\BibitemShut {NoStop}%
\bibitem [{\citenamefont {Jander}\ \emph {et~al.}(2001)\citenamefont {Jander},
  \citenamefont {Moreland},\ and\ \citenamefont {Kabos}}]{Jander2001}%
  \BibitemOpen
  \bibfield  {author} {\bibinfo {author} {\bibfnamefont {A.}~\bibnamefont
  {Jander}}, \bibinfo {author} {\bibfnamefont {J.}~\bibnamefont {Moreland}}, \
  and\ \bibinfo {author} {\bibfnamefont {P.}~\bibnamefont {Kabos}},\ }\href
  {\doibase 10.1063/1.1361095} {\bibfield  {journal} {\bibinfo  {journal}
  {Appl. Phys. Lett}\ }\textbf {\bibinfo {volume} {78}},\ \bibinfo {pages}
  {2348} (\bibinfo {year} {2001})}\BibitemShut {NoStop}%
\bibitem [{\citenamefont {Losby}\ \emph {et~al.}(2018)\citenamefont {Losby},
  \citenamefont {Sauer},\ and\ \citenamefont {Freeman}}]{Losby2018}%
  \BibitemOpen
  \bibfield  {author} {\bibinfo {author} {\bibfnamefont {J.~E.}\ \bibnamefont
  {Losby}}, \bibinfo {author} {\bibfnamefont {V.~T.~K.}\ \bibnamefont {Sauer}},
  \ and\ \bibinfo {author} {\bibfnamefont {M.~R.}\ \bibnamefont {Freeman}},\
  }\href {https://iopscience.iop.org/article/10.1088/1361-6463/aadccb}
  {\bibfield  {journal} {\bibinfo  {journal} {J. Phys. D: Appl. Phys}\ }\textbf
  {\bibinfo {volume} {51}},\ \bibinfo {pages} {483001} (\bibinfo {year}
  {2018})}\BibitemShut {NoStop}%
\bibitem [{\citenamefont {Harii}\ \emph {et~al.}(2019)\citenamefont {Harii},
  \citenamefont {Seo}, \citenamefont {Tsutsumi}, \citenamefont {Chudo},
  \citenamefont {Oyanagi}, \citenamefont {Matsuo}, \citenamefont {Shiomi},
  \citenamefont {Ono}, \citenamefont {Maekawa},\ and\ \citenamefont
  {Saitoh}}]{Harii2019}%
  \BibitemOpen
  \bibfield  {author} {\bibinfo {author} {\bibfnamefont {K.}~\bibnamefont
  {Harii}}, \bibinfo {author} {\bibfnamefont {Y.-J.}\ \bibnamefont {Seo}},
  \bibinfo {author} {\bibfnamefont {Y.}~\bibnamefont {Tsutsumi}}, \bibinfo
  {author} {\bibfnamefont {H.}~\bibnamefont {Chudo}}, \bibinfo {author}
  {\bibfnamefont {K.}~\bibnamefont {Oyanagi}}, \bibinfo {author} {\bibfnamefont
  {M.}~\bibnamefont {Matsuo}}, \bibinfo {author} {\bibfnamefont
  {Y.}~\bibnamefont {Shiomi}}, \bibinfo {author} {\bibfnamefont
  {T.}~\bibnamefont {Ono}}, \bibinfo {author} {\bibfnamefont {S.}~\bibnamefont
  {Maekawa}}, \ and\ \bibinfo {author} {\bibfnamefont {E.}~\bibnamefont
  {Saitoh}},\ }\href {https://www.nature.com/articles/s41467-019-10625-y}
  {\bibfield  {journal} {\bibinfo  {journal} {Nat. Comms.}\ }\textbf {\bibinfo
  {volume} {10}},\ \bibinfo {pages} {2616} (\bibinfo {year}
  {2019})}\BibitemShut {NoStop}%
\bibitem [{\citenamefont {Eichenfield}\ \emph {et~al.}(2009)\citenamefont
  {Eichenfield}, \citenamefont {Camacho}, \citenamefont {Chan}, \citenamefont
  {Vahala},\ and\ \citenamefont {Painter}}]{Eichenfield2009}%
  \BibitemOpen
  \bibfield  {author} {\bibinfo {author} {\bibfnamefont {M.}~\bibnamefont
  {Eichenfield}}, \bibinfo {author} {\bibfnamefont {R.}~\bibnamefont
  {Camacho}}, \bibinfo {author} {\bibfnamefont {J.}~\bibnamefont {Chan}},
  \bibinfo {author} {\bibfnamefont {K.~J.}\ \bibnamefont {Vahala}}, \ and\
  \bibinfo {author} {\bibfnamefont {O.}~\bibnamefont {Painter}},\ }\href
  {\doibase 10.1038/nature08061} {\bibfield  {journal} {\bibinfo  {journal}
  {Nature}\ }\textbf {\bibinfo {volume} {459}},\ \bibinfo {pages} {550}
  (\bibinfo {year} {2009})}\BibitemShut {NoStop}%
\bibitem [{\citenamefont {Kim}\ \emph {et~al.}(2013)\citenamefont {Kim},
  \citenamefont {Doolin}, \citenamefont {Hauer}, \citenamefont {MacDonald},
  \citenamefont {Freeman}, \citenamefont {Barclay},\ and\ \citenamefont
  {Davis}}]{Kim2013}%
  \BibitemOpen
  \bibfield  {author} {\bibinfo {author} {\bibfnamefont {P.~H.}\ \bibnamefont
  {Kim}}, \bibinfo {author} {\bibfnamefont {C.}~\bibnamefont {Doolin}},
  \bibinfo {author} {\bibfnamefont {B.~D.}\ \bibnamefont {Hauer}}, \bibinfo
  {author} {\bibfnamefont {A.~J.}\ \bibnamefont {MacDonald}}, \bibinfo {author}
  {\bibfnamefont {M.~R.}\ \bibnamefont {Freeman}}, \bibinfo {author}
  {\bibfnamefont {P.~E.}\ \bibnamefont {Barclay}}, \ and\ \bibinfo {author}
  {\bibfnamefont {J.~P.}\ \bibnamefont {Davis}},\ }\href {\doibase
  10.1063/1.4789442} {\bibfield  {journal} {\bibinfo  {journal} {Applied
  Physics Letters}\ }\textbf {\bibinfo {volume} {102}},\ \bibinfo {pages}
  {053102} (\bibinfo {year} {2013})},\ \Eprint {http://arxiv.org/abs/1210.1852}
  {1210.1852} \BibitemShut {NoStop}%
\bibitem [{\citenamefont {Aspelmeyer}\ \emph {et~al.}(2014)\citenamefont
  {Aspelmeyer}, \citenamefont {Kippenberg},\ and\ \citenamefont
  {Marquardt}}]{Aspelmeyer2014}%
  \BibitemOpen
  \bibfield  {author} {\bibinfo {author} {\bibfnamefont {M.}~\bibnamefont
  {Aspelmeyer}}, \bibinfo {author} {\bibfnamefont {T.~J.}\ \bibnamefont
  {Kippenberg}}, \ and\ \bibinfo {author} {\bibfnamefont {F.}~\bibnamefont
  {Marquardt}},\ }\href {\doibase 10.1103/RevModPhys.86.1391} {\bibfield
  {journal} {\bibinfo  {journal} {Reviews of Modern Physics}\ }\textbf
  {\bibinfo {volume} {86}},\ \bibinfo {pages} {1391} (\bibinfo {year}
  {2014})}\BibitemShut {NoStop}%
\bibitem [{\citenamefont {Forstner}\ \emph {et~al.}(2014)\citenamefont
  {Forstner}, \citenamefont {Sheridan}, \citenamefont {Knittel}, \citenamefont
  {Humphreys}, \citenamefont {Brawley}, \citenamefont {Rubinsztein-Dunlop},\
  and\ \citenamefont {Bowen}}]{Forstner2014}%
  \BibitemOpen
  \bibfield  {author} {\bibinfo {author} {\bibfnamefont {S.}~\bibnamefont
  {Forstner}}, \bibinfo {author} {\bibfnamefont {E.}~\bibnamefont {Sheridan}},
  \bibinfo {author} {\bibfnamefont {J.}~\bibnamefont {Knittel}}, \bibinfo
  {author} {\bibfnamefont {C.~L.}\ \bibnamefont {Humphreys}}, \bibinfo {author}
  {\bibfnamefont {G.~A.}\ \bibnamefont {Brawley}}, \bibinfo {author}
  {\bibfnamefont {H.}~\bibnamefont {Rubinsztein-Dunlop}}, \ and\ \bibinfo
  {author} {\bibfnamefont {W.~P.}\ \bibnamefont {Bowen}},\ }\href {\doibase
  10.1002/adma.201401144} {\bibfield  {journal} {\bibinfo  {journal} {Advanced
  Materials}\ }\textbf {\bibinfo {volume} {26}},\ \bibinfo {pages} {6348}
  (\bibinfo {year} {2014})}\BibitemShut {NoStop}%
\bibitem [{\citenamefont {Li}\ \emph {et~al.}(2020)\citenamefont {Li},
  \citenamefont {Brawley}, \citenamefont {Greenall}, \citenamefont {Forstner},
  \citenamefont {Sheridan}, \citenamefont {Rubinsztein-Dunlop},\ and\
  \citenamefont {Bowen}}]{Li2020}%
  \BibitemOpen
  \bibfield  {author} {\bibinfo {author} {\bibfnamefont {B.-B.}\ \bibnamefont
  {Li}}, \bibinfo {author} {\bibfnamefont {G.}~\bibnamefont {Brawley}},
  \bibinfo {author} {\bibfnamefont {H.}~\bibnamefont {Greenall}}, \bibinfo
  {author} {\bibfnamefont {S.}~\bibnamefont {Forstner}}, \bibinfo {author}
  {\bibfnamefont {E.}~\bibnamefont {Sheridan}}, \bibinfo {author}
  {\bibfnamefont {H.}~\bibnamefont {Rubinsztein-Dunlop}}, \ and\ \bibinfo
  {author} {\bibfnamefont {W.~P.}\ \bibnamefont {Bowen}},\ }\href {\doibase
  10.1364/prj.390261} {\bibfield  {journal} {\bibinfo  {journal} {Photonics
  Research}\ }\textbf {\bibinfo {volume} {8}},\ \bibinfo {pages} {1064}
  (\bibinfo {year} {2020})},\ \Eprint {http://arxiv.org/abs/2002.00728}
  {2002.00728} \BibitemShut {NoStop}%
\bibitem [{\citenamefont {Kovalev}\ \emph {et~al.}(2005)\citenamefont
  {Kovalev}, \citenamefont {Bauer},\ and\ \citenamefont
  {Brataas}}]{Kovalev2005}%
  \BibitemOpen
  \bibfield  {author} {\bibinfo {author} {\bibfnamefont {A.~A.}\ \bibnamefont
  {Kovalev}}, \bibinfo {author} {\bibfnamefont {G.~E.~W.}\ \bibnamefont
  {Bauer}}, \ and\ \bibinfo {author} {\bibfnamefont {A.}~\bibnamefont
  {Brataas}},\ }\href {\doibase 10.1103/PhysRevLett.94.167201} {\bibfield
  {journal} {\bibinfo  {journal} {Phys. Rev. Lett.}\ }\textbf {\bibinfo
  {volume} {94}},\ \bibinfo {pages} {167201} (\bibinfo {year}
  {2005})}\BibitemShut {NoStop}%
\bibitem [{\citenamefont {Kovalev}\ \emph {et~al.}(2006)\citenamefont
  {Kovalev}, \citenamefont {Bauer},\ and\ \citenamefont
  {Brataas}}]{Kovalev2006}%
  \BibitemOpen
  \bibfield  {author} {\bibinfo {author} {\bibfnamefont {A.~A.}\ \bibnamefont
  {Kovalev}}, \bibinfo {author} {\bibfnamefont {G.~E.~W.}\ \bibnamefont
  {Bauer}}, \ and\ \bibinfo {author} {\bibfnamefont {A.}~\bibnamefont
  {Brataas}},\ }\href {\doibase 10.1143/JJAP.45.3878} {\bibfield  {journal}
  {\bibinfo  {journal} {Jpn. J. Appl. Phys.}\ }\textbf {\bibinfo {volume}
  {45}},\ \bibinfo {pages} {3878} (\bibinfo {year} {2006})}\BibitemShut
  {NoStop}%
\bibitem [{\citenamefont {Wu}\ \emph {et~al.}(2017)\citenamefont {Wu},
  \citenamefont {Wu}, \citenamefont {Firdous}, \citenamefont {Fani~Sani},
  \citenamefont {Losby}, \citenamefont {Freeman},\ and\ \citenamefont
  {Barclay}}]{Wu2017}%
  \BibitemOpen
  \bibfield  {author} {\bibinfo {author} {\bibfnamefont {M.}~\bibnamefont
  {Wu}}, \bibinfo {author} {\bibfnamefont {N.~L.~Y.}\ \bibnamefont {Wu}},
  \bibinfo {author} {\bibfnamefont {T.}~\bibnamefont {Firdous}}, \bibinfo
  {author} {\bibfnamefont {F.}~\bibnamefont {Fani~Sani}}, \bibinfo {author}
  {\bibfnamefont {J.~E.}\ \bibnamefont {Losby}}, \bibinfo {author}
  {\bibfnamefont {M.~R.}\ \bibnamefont {Freeman}}, \ and\ \bibinfo {author}
  {\bibfnamefont {P.~E.}\ \bibnamefont {Barclay}},\ }\href {\doibase
  10.1038/NNANO.2016.226} {\bibfield  {journal} {\bibinfo  {journal} {Nat.
  Nanotechnol.}\ }\textbf {\bibinfo {volume} {12}},\ \bibinfo {pages} {127}
  (\bibinfo {year} {2017})}\BibitemShut {NoStop}%
\bibitem [{\citenamefont {Hajisalem}\ \emph {et~al.}(2019)\citenamefont
  {Hajisalem}, \citenamefont {Losby}, \citenamefont {de~Oliveira~Luiz},
  \citenamefont {Sauer}, \citenamefont {Barclay},\ and\ \citenamefont
  {Freeman}}]{Hajisalem2019}%
  \BibitemOpen
  \bibfield  {author} {\bibinfo {author} {\bibfnamefont {G.}~\bibnamefont
  {Hajisalem}}, \bibinfo {author} {\bibfnamefont {J.~E.}\ \bibnamefont
  {Losby}}, \bibinfo {author} {\bibfnamefont {G.}~\bibnamefont
  {de~Oliveira~Luiz}}, \bibinfo {author} {\bibfnamefont {V.~T.~K.}\
  \bibnamefont {Sauer}}, \bibinfo {author} {\bibfnamefont {P.~E.}\ \bibnamefont
  {Barclay}}, \ and\ \bibinfo {author} {\bibfnamefont {M.~R.}\ \bibnamefont
  {Freeman}},\ }\href {\doibase 10.1088/1367-2630/ab4386} {\bibfield  {journal}
  {\bibinfo  {journal} {New J. Phys.}\ }\textbf {\bibinfo {volume} {21}},\
  \bibinfo {pages} {095005} (\bibinfo {year} {2019})}\BibitemShut {NoStop}%
\bibitem [{\citenamefont {Hryciw}\ and\ \citenamefont
  {Barclay}(2013)}]{Hryciw2013}%
  \BibitemOpen
  \bibfield  {author} {\bibinfo {author} {\bibfnamefont {A.~C.}\ \bibnamefont
  {Hryciw}}\ and\ \bibinfo {author} {\bibfnamefont {P.~E.}\ \bibnamefont
  {Barclay}},\ }\href {https://doi.org/10.1364/OL.38.001612} {\bibfield
  {journal} {\bibinfo  {journal} {Opt. Lett.}\ }\textbf {\bibinfo {volume}
  {38}},\ \bibinfo {pages} {1612} (\bibinfo {year} {2013})}\BibitemShut
  {NoStop}%
\bibitem [{COM()}]{COMSOL}%
  \BibitemOpen
  \href {https://comsol.com/} {\enquote {\bibinfo {title} {{COMSOL
  Multiphysics\textsuperscript{\textregistered} v 6.1. www.comsol.com. COMSOL
  AB, Stockholm, Sweden.}}}\ }\BibitemShut {NoStop}%
\bibitem [{See Supplemental Information at [URL] for further information on
  data processing techniques, macrospin simulation details, and Mumax$^3$
  simulation details()}]{Supplement}%
  \BibitemOpen
  See Supplemental Information at [URL] for further information on data
  processing techniques, macrospin simulation details, and Mumax$^3$ simulation
  details,\ \href@noop {} {}\BibitemShut {NoStop}%
\bibitem [{\citenamefont {Sang-Hyun}\ \emph {et~al.}(2014)\citenamefont
  {Sang-Hyun}, \citenamefont {Imtiaz}, \citenamefont {Wallis}, \citenamefont
  {Russek}, \citenamefont {Kabos}, \citenamefont {Cai},\ and\ \citenamefont
  {Chudnovsky}}]{Lim2014}%
  \BibitemOpen
  \bibfield  {author} {\bibinfo {author} {\bibfnamefont {L.}~\bibnamefont
  {Sang-Hyun}}, \bibinfo {author} {\bibfnamefont {A.}~\bibnamefont {Imtiaz}},
  \bibinfo {author} {\bibfnamefont {T.~M.}\ \bibnamefont {Wallis}}, \bibinfo
  {author} {\bibfnamefont {S.}~\bibnamefont {Russek}}, \bibinfo {author}
  {\bibfnamefont {P.}~\bibnamefont {Kabos}}, \bibinfo {author} {\bibfnamefont
  {L.}~\bibnamefont {Cai}}, \ and\ \bibinfo {author} {\bibfnamefont {E.~M.}\
  \bibnamefont {Chudnovsky}},\ }\href {\doibase 10.1209/0295-5075/105/37009}
  {\bibfield  {journal} {\bibinfo  {journal} {EPL}\ }\textbf {\bibinfo {volume}
  {105}},\ \bibinfo {pages} {37009} (\bibinfo {year} {2014})}\BibitemShut
  {NoStop}%
\bibitem [{\citenamefont {Cowburn}\ \emph {et~al.}(1999)\citenamefont
  {Cowburn}, \citenamefont {Koltsov}, \citenamefont {Adeyeye}, \citenamefont
  {Welland},\ and\ \citenamefont {Tricker}}]{Cowburn1999}%
  \BibitemOpen
  \bibfield  {author} {\bibinfo {author} {\bibfnamefont {R.~P.}\ \bibnamefont
  {Cowburn}}, \bibinfo {author} {\bibfnamefont {D.~K.}\ \bibnamefont
  {Koltsov}}, \bibinfo {author} {\bibfnamefont {A.~O.}\ \bibnamefont
  {Adeyeye}}, \bibinfo {author} {\bibfnamefont {M.~E.}\ \bibnamefont
  {Welland}}, \ and\ \bibinfo {author} {\bibfnamefont {D.~M.}\ \bibnamefont
  {Tricker}},\ }\href {\doibase 10.1103/PhysRevLett.83.1042} {\bibfield
  {journal} {\bibinfo  {journal} {Phys. Rev. Lett.}\ }\textbf {\bibinfo
  {volume} {83}},\ \bibinfo {pages} {1042} (\bibinfo {year}
  {1999})}\BibitemShut {NoStop}%
\bibitem [{\citenamefont {Burgess}\ \emph {et~al.}(2013)\citenamefont
  {Burgess}, \citenamefont {Fraser}, \citenamefont {Fani~Sani}, \citenamefont
  {Vick}, \citenamefont {Hauer}, \citenamefont {Davis},\ and\ \citenamefont
  {Freeman}}]{Burgess2013}%
  \BibitemOpen
  \bibfield  {author} {\bibinfo {author} {\bibfnamefont {J.~A.~J.}\
  \bibnamefont {Burgess}}, \bibinfo {author} {\bibfnamefont {A.~E.}\
  \bibnamefont {Fraser}}, \bibinfo {author} {\bibfnamefont {F.}~\bibnamefont
  {Fani~Sani}}, \bibinfo {author} {\bibfnamefont {D.}~\bibnamefont {Vick}},
  \bibinfo {author} {\bibfnamefont {B.~D.}\ \bibnamefont {Hauer}}, \bibinfo
  {author} {\bibfnamefont {J.~P.}\ \bibnamefont {Davis}}, \ and\ \bibinfo
  {author} {\bibfnamefont {M.~R.}\ \bibnamefont {Freeman}},\ }\href {\doibase
  10.1126/science.1231390} {\bibfield  {journal} {\bibinfo  {journal}
  {Science}\ }\textbf {\bibinfo {volume} {339}},\ \bibinfo {pages} {1051}
  (\bibinfo {year} {2013})}\BibitemShut {NoStop}%
\bibitem [{\citenamefont {Compton}\ and\ \citenamefont
  {Crowell}(2006)}]{Compton2006}%
  \BibitemOpen
  \bibfield  {author} {\bibinfo {author} {\bibfnamefont {R.~L.}\ \bibnamefont
  {Compton}}\ and\ \bibinfo {author} {\bibfnamefont {P.~A.}\ \bibnamefont
  {Crowell}},\ }\href {\doibase 10.1103/PhysRevLett.97.137202} {\bibfield
  {journal} {\bibinfo  {journal} {Phys. Rev. Lett.}\ }\textbf {\bibinfo
  {volume} {97}},\ \bibinfo {pages} {137202} (\bibinfo {year}
  {2006})}\BibitemShut {NoStop}%
\bibitem [{\citenamefont {Compton}\ \emph {et~al.}(2010)\citenamefont
  {Compton}, \citenamefont {Chen},\ and\ \citenamefont
  {Crowell}}]{Compton2010}%
  \BibitemOpen
  \bibfield  {author} {\bibinfo {author} {\bibfnamefont {R.~L.}\ \bibnamefont
  {Compton}}, \bibinfo {author} {\bibfnamefont {T.~Y.}\ \bibnamefont {Chen}}, \
  and\ \bibinfo {author} {\bibfnamefont {P.~A.}\ \bibnamefont {Crowell}},\
  }\href {\doibase 10.1103/PhysRevB.81.144412} {\bibfield  {journal} {\bibinfo
  {journal} {Phys. Rev. B}\ }\textbf {\bibinfo {volume} {81}},\ \bibinfo
  {pages} {144412} (\bibinfo {year} {2010})}\BibitemShut {NoStop}%
\bibitem [{\citenamefont {Vansteenkiste}\ \emph {et~al.}(2014)\citenamefont
  {Vansteenkiste}, \citenamefont {Leliaert}, \citenamefont {Dvornik},
  \citenamefont {Helsen}, \citenamefont {Garcia-Sanchez},\ and\ \citenamefont
  {Van~de Wiele}}]{Vansteenkiste2014}%
  \BibitemOpen
  \bibfield  {author} {\bibinfo {author} {\bibfnamefont {A.}~\bibnamefont
  {Vansteenkiste}}, \bibinfo {author} {\bibfnamefont {J.}~\bibnamefont
  {Leliaert}}, \bibinfo {author} {\bibfnamefont {M.}~\bibnamefont {Dvornik}},
  \bibinfo {author} {\bibfnamefont {M.}~\bibnamefont {Helsen}}, \bibinfo
  {author} {\bibfnamefont {F.}~\bibnamefont {Garcia-Sanchez}}, \ and\ \bibinfo
  {author} {\bibfnamefont {B.}~\bibnamefont {Van~de Wiele}},\ }\href@noop {}
  {\bibfield  {journal} {\bibinfo  {journal} {AIP Adv.}\ }\textbf {\bibinfo
  {volume} {4}},\ \bibinfo {pages} {107133} (\bibinfo {year}
  {2014})}\BibitemShut {NoStop}%
\bibitem [{\citenamefont {Dornes}\ \emph {et~al.}(2019)\citenamefont {Dornes},
  \citenamefont {Acremann}, \citenamefont {Savoini}, \citenamefont {Kubli},
  \citenamefont {Neugebauer}, \citenamefont {Abreu}, \citenamefont {Huber},
  \citenamefont {Lantz}, \citenamefont {Vaz}, \citenamefont {Lemke},
  \citenamefont {Bothschafter}, \citenamefont {Porer}, \citenamefont
  {Esposito}, \citenamefont {Rettig}, \citenamefont {Buzzi}, \citenamefont
  {Alberca}, \citenamefont {Windsor}, \citenamefont {Beaud}, \citenamefont
  {Staub}, \citenamefont {Zhu}, \citenamefont {Song}, \citenamefont {Glownia},\
  and\ \citenamefont {Johnson}}]{Dornes2019}%
  \BibitemOpen
  \bibfield  {author} {\bibinfo {author} {\bibfnamefont {C.}~\bibnamefont
  {Dornes}}, \bibinfo {author} {\bibfnamefont {Y.}~\bibnamefont {Acremann}},
  \bibinfo {author} {\bibfnamefont {M.}~\bibnamefont {Savoini}}, \bibinfo
  {author} {\bibfnamefont {M.}~\bibnamefont {Kubli}}, \bibinfo {author}
  {\bibfnamefont {M.~J.}\ \bibnamefont {Neugebauer}}, \bibinfo {author}
  {\bibfnamefont {E.}~\bibnamefont {Abreu}}, \bibinfo {author} {\bibfnamefont
  {L.}~\bibnamefont {Huber}}, \bibinfo {author} {\bibfnamefont
  {G.}~\bibnamefont {Lantz}}, \bibinfo {author} {\bibfnamefont {C.~A.~F.}\
  \bibnamefont {Vaz}}, \bibinfo {author} {\bibfnamefont {H.}~\bibnamefont
  {Lemke}}, \bibinfo {author} {\bibfnamefont {E.~M.}\ \bibnamefont
  {Bothschafter}}, \bibinfo {author} {\bibfnamefont {M.}~\bibnamefont {Porer}},
  \bibinfo {author} {\bibfnamefont {V.}~\bibnamefont {Esposito}}, \bibinfo
  {author} {\bibfnamefont {L.}~\bibnamefont {Rettig}}, \bibinfo {author}
  {\bibfnamefont {M.}~\bibnamefont {Buzzi}}, \bibinfo {author} {\bibfnamefont
  {A.}~\bibnamefont {Alberca}}, \bibinfo {author} {\bibfnamefont {Y.~W.}\
  \bibnamefont {Windsor}}, \bibinfo {author} {\bibfnamefont {P.}~\bibnamefont
  {Beaud}}, \bibinfo {author} {\bibfnamefont {U.}~\bibnamefont {Staub}},
  \bibinfo {author} {\bibfnamefont {D.}~\bibnamefont {Zhu}}, \bibinfo {author}
  {\bibfnamefont {S.}~\bibnamefont {Song}}, \bibinfo {author} {\bibfnamefont
  {J.~M.}\ \bibnamefont {Glownia}}, \ and\ \bibinfo {author} {\bibfnamefont
  {S.~L.}\ \bibnamefont {Johnson}},\ }\href {\doibase
  https://doi.org/10.1038/s41586-018-0822-7} {\bibfield  {journal} {\bibinfo
  {journal} {Nature}\ }\textbf {\bibinfo {volume} {565}},\ \bibinfo {pages}
  {209} (\bibinfo {year} {2019})}\BibitemShut {NoStop}%
\end{thebibliography}

\begin{thebibliography}{6}%
\makeatletter
\providecommand \@ifxundefined [1]{%
 \@ifx{#1\undefined}
}%
\providecommand \@ifnum [1]{%
 \ifnum #1\expandafter \@firstoftwo
 \else \expandafter \@secondoftwo
 \fi
}%
\providecommand \@ifx [1]{%
 \ifx #1\expandafter \@firstoftwo
 \else \expandafter \@secondoftwo
 \fi
}%
\providecommand \natexlab [1]{#1}%
\providecommand \enquote  [1]{``#1''}%
\providecommand \bibnamefont  [1]{#1}%
\providecommand \bibfnamefont [1]{#1}%
\providecommand \citenamefont [1]{#1}%
\providecommand \href@noop [0]{\@secondoftwo}%
\providecommand \href [0]{\begingroup \@sanitize@url \@href}%
\providecommand \@href[1]{\@@startlink{#1}\@@href}%
\providecommand \@@href[1]{\endgroup#1\@@endlink}%
\providecommand \@sanitize@url [0]{\catcode `\\12\catcode `\$12\catcode
  `\&12\catcode `\#12\catcode `\^12\catcode `\_12\catcode `\%12\relax}%
\providecommand \@@startlink[1]{}%
\providecommand \@@endlink[0]{}%
\providecommand \url  [0]{\begingroup\@sanitize@url \@url }%
\providecommand \@url [1]{\endgroup\@href {#1}{\urlprefix }}%
\providecommand \urlprefix  [0]{URL }%
\providecommand \Eprint [0]{\href }%
\providecommand \doibase [0]{http://dx.doi.org/}%
\providecommand \selectlanguage [0]{\@gobble}%
\providecommand \bibinfo  [0]{\@secondoftwo}%
\providecommand \bibfield  [0]{\@secondoftwo}%
\providecommand \translation [1]{[#1]}%
\providecommand \BibitemOpen [0]{}%
\providecommand \bibitemStop [0]{}%
\providecommand \bibitemNoStop [0]{.\EOS\space}%
\providecommand \EOS [0]{\spacefactor3000\relax}%
\providecommand \BibitemShut  [1]{\csname bibitem#1\endcsname}%
\let\auto@bib@innerbib\@empty
%</preamble>
\bibitem [{\citenamefont {Cleland}(2003)}]{Cleland2003}%
  \BibitemOpen
  \bibfield  {author} {\bibinfo {author} {\bibfnamefont {A.~N.}\ \bibnamefont
  {Cleland}},\ }\href@noop {} {\emph {\bibinfo {title} {Foundations of
  nanomechanics : from solid-state theory to device applications.}}},\ Advanced
  texts in physics\ (\bibinfo  {publisher} {Springer},\ \bibinfo {year}
  {2003})\BibitemShut {NoStop}%
\bibitem [{\citenamefont {Fast}\ and\ \citenamefont
  {Freeman}(2023)}]{GitHub_Macrospin}%
  \BibitemOpen
  \bibfield  {author} {\bibinfo {author} {\bibfnamefont {K.}~\bibnamefont
  {Fast}}\ and\ \bibinfo {author} {\bibfnamefont {M.}~\bibnamefont {Freeman}},\
  }\href@noop {} {\enquote {\bibinfo {title} {{MacrospinSimulation}},}\ }
  (\bibinfo {year} {2023})\BibitemShut {NoStop}%
\bibitem [{\citenamefont {Mori}\ \emph {et~al.}(2020)\citenamefont {Mori},
  \citenamefont {Dunsmore}, \citenamefont {Losby}, \citenamefont {Jenson},
  \citenamefont {Belov},\ and\ \citenamefont {Freeman}}]{Mori2020}%
  \BibitemOpen
  \bibfield  {author} {\bibinfo {author} {\bibfnamefont {K.}~\bibnamefont
  {Mori}}, \bibinfo {author} {\bibfnamefont {M.~G.}\ \bibnamefont {Dunsmore}},
  \bibinfo {author} {\bibfnamefont {J.~E.}\ \bibnamefont {Losby}}, \bibinfo
  {author} {\bibfnamefont {D.~M.}\ \bibnamefont {Jenson}}, \bibinfo {author}
  {\bibfnamefont {M.}~\bibnamefont {Belov}}, \ and\ \bibinfo {author}
  {\bibfnamefont {M.~R.}\ \bibnamefont {Freeman}},\ }\href
  {https://doi.org/10.1103/PhysRevB.102.054415} {\bibfield  {journal} {\bibinfo
   {journal} {Phys. Rev. B}\ }\textbf {\bibinfo {volume} {102}},\ \bibinfo
  {pages} {054415} (\bibinfo {year} {2020})}\BibitemShut {NoStop}%
\bibitem [{\citenamefont {Vansteenkiste}\ \emph {et~al.}(2014)\citenamefont
  {Vansteenkiste}, \citenamefont {Leliaert}, \citenamefont {Dvornik},
  \citenamefont {Helsen}, \citenamefont {Garcia-Sanchez},\ and\ \citenamefont
  {Van~de Wiele}}]{Vansteenkiste2014}%
  \BibitemOpen
  \bibfield  {author} {\bibinfo {author} {\bibfnamefont {A.}~\bibnamefont
  {Vansteenkiste}}, \bibinfo {author} {\bibfnamefont {J.}~\bibnamefont
  {Leliaert}}, \bibinfo {author} {\bibfnamefont {M.}~\bibnamefont {Dvornik}},
  \bibinfo {author} {\bibfnamefont {M.}~\bibnamefont {Helsen}}, \bibinfo
  {author} {\bibfnamefont {F.}~\bibnamefont {Garcia-Sanchez}}, \ and\ \bibinfo
  {author} {\bibfnamefont {B.}~\bibnamefont {Van~de Wiele}},\ }\href@noop {}
  {\bibfield  {journal} {\bibinfo  {journal} {AIP Adv.}\ }\textbf {\bibinfo
  {volume} {4}},\ \bibinfo {pages} {107133} (\bibinfo {year}
  {2014})}\BibitemShut {NoStop}%
\bibitem [{\citenamefont {Losby}\ \emph {et~al.}(2015)\citenamefont {Losby},
  \citenamefont {Sani}, \citenamefont {Grandmont}, \citenamefont {Diao},
  \citenamefont {Belov}, \citenamefont {Burgess}, \citenamefont {Compton},
  \citenamefont {Hiebert}, \citenamefont {Vick}, \citenamefont {Mohammad},
  \citenamefont {Salimi}, \citenamefont {Bridges}, \citenamefont {Thomson},\
  and\ \citenamefont {Freeman}}]{Losby2015}%
  \BibitemOpen
  \bibfield  {author} {\bibinfo {author} {\bibfnamefont {J.~E.}\ \bibnamefont
  {Losby}}, \bibinfo {author} {\bibfnamefont {F.~F.}\ \bibnamefont {Sani}},
  \bibinfo {author} {\bibfnamefont {D.~T.}\ \bibnamefont {Grandmont}}, \bibinfo
  {author} {\bibfnamefont {Z.}~\bibnamefont {Diao}}, \bibinfo {author}
  {\bibfnamefont {M.}~\bibnamefont {Belov}}, \bibinfo {author} {\bibfnamefont
  {J.~A.~J.}\ \bibnamefont {Burgess}}, \bibinfo {author} {\bibfnamefont
  {S.~R.}\ \bibnamefont {Compton}}, \bibinfo {author} {\bibfnamefont {W.~K.}\
  \bibnamefont {Hiebert}}, \bibinfo {author} {\bibfnamefont {D.}~\bibnamefont
  {Vick}}, \bibinfo {author} {\bibfnamefont {K.}~\bibnamefont {Mohammad}},
  \bibinfo {author} {\bibfnamefont {E.}~\bibnamefont {Salimi}}, \bibinfo
  {author} {\bibfnamefont {G.~E.}\ \bibnamefont {Bridges}}, \bibinfo {author}
  {\bibfnamefont {D.~J.}\ \bibnamefont {Thomson}}, \ and\ \bibinfo {author}
  {\bibfnamefont {M.~R.}\ \bibnamefont {Freeman}},\ }\href {\doibase
  10.1126/science.aad2449} {\bibfield  {journal} {\bibinfo  {journal}
  {Science}\ }\textbf {\bibinfo {volume} {350}},\ \bibinfo {pages} {798}
  (\bibinfo {year} {2015})}\BibitemShut {NoStop}%
\bibitem [{\citenamefont {Hajisalem}\ \emph {et~al.}(2019)\citenamefont
  {Hajisalem}, \citenamefont {Losby}, \citenamefont {de~Oliveira~Luiz},
  \citenamefont {Sauer}, \citenamefont {Barclay},\ and\ \citenamefont
  {Freeman}}]{Hajisalem2019}%
  \BibitemOpen
  \bibfield  {author} {\bibinfo {author} {\bibfnamefont {G.}~\bibnamefont
  {Hajisalem}}, \bibinfo {author} {\bibfnamefont {J.~E.}\ \bibnamefont
  {Losby}}, \bibinfo {author} {\bibfnamefont {G.}~\bibnamefont
  {de~Oliveira~Luiz}}, \bibinfo {author} {\bibfnamefont {V.~T.~K.}\
  \bibnamefont {Sauer}}, \bibinfo {author} {\bibfnamefont {P.~E.}\ \bibnamefont
  {Barclay}}, \ and\ \bibinfo {author} {\bibfnamefont {M.~R.}\ \bibnamefont
  {Freeman}},\ }\href {\doibase 10.1088/1367-2630/ab4386} {\bibfield  {journal}
  {\bibinfo  {journal} {New J. Phys.}\ }\textbf {\bibinfo {volume} {21}},\
  \bibinfo {pages} {095005} (\bibinfo {year} {2019})}\BibitemShut {NoStop}%
\end{thebibliography}

\newpage
\beginsupplement
%\chapter{Einstein-de Haas torque as a discrete spectroscopic probe allows nanomechanical measurement of a magnetic resonance:\\Supplementary Information}

%\begin{document}
\begin{center}
\textbf{Einstein-de Haas torque as a discrete spectroscopic probe allows nanomechanical measurement of a magnetic resonance:\\Supplementary Information}\\
K.R. Fast$^{\mathrm{1,3,*}}$, J.E. Losby$^{\mathrm{2,3,*}}$, G. Hajisalem$^{\mathrm{2,3}}$, P.E. Barclay$^{\mathrm{2,3}}$, M.R. Freeman$^{\mathrm{1,3}}$\\
\textit{1. Department of Physics,\\University of Alberta, Edmonton, Alberta T6G 2E1, Canada}\\
\textit{2. Department of Physics and Astronomy,\\University of Calgary, Calgary, Alberta T2N 1N4, Canada}\\
\textit{3. Nanotechnology Research Centre,\\National Research Council of Canada,\\Edmonton, Alberta T6G 2M9, Canada}
%\author{K.R. Fast$^{1,3,^{\ast}}$, J.E. Losby$^{2,3,^{\ast}}$, G. Hajisalem$^{2,3}$, P.E. Barclay$^{2,3}$, M.R. Freeman$^{1,3}$}
%
%\affiliation{1. 
%Department of Physics, University of Alberta, Edmonton, Alberta T6G 2E1, Canada%\\
%}
%\affiliation{2. 
%Department of Physics and Astronomy, University of Calgary, Calgary, Alberta T2N 1N4, Canada%\\
%}
%\affiliation{3.
%Nanotechnology Research Centre, National Research Council
%of Canada, Edmonton, Alberta T6G 2M9, Canada%\\
%}
%
%\email{Both authors contributed equally to this work. email: pbarclay@ucalgary.ca , freemanm@ualberta.ca}
%
%\maketitle
\end{center}
\section{Mechanical Resonance Frequencies}
\subsection{Euler-Bernoulli Mode Frequencies} \label{suppSect:EBFrequencies}
The ladder of mechanical modes driven by $y$-torques closely match flexural modes predicted by Euler-Bernoulli beam theory for a free-free beam. The equation defining the eigenmodes for a beam of length $L$ with free ends is \cite{Cleland2003}
\begin{equation}
    \cos{\beta_nL}\cosh{\beta_nL} = 1.
    \label{suppEq:EBtheory}
\end{equation}
 The roots of \autoref{suppEq:EBtheory}, $\beta_n$, determine the flexural mode frequencies, $\omega_n = \beta_n^2 \sqrt{\frac{EI}{M/L}}$, for a beam with elastic modulus $E$, moment of area $I$, and mass $M$. For out-of-plane rotation about the $y$-axis, the moment of area for a beam with thickness $t$ and width $w$ is $I = \frac{wt^3}{12}$. The sensor is approximated to be a uniform beam of silicon with elastic modulus 170 GPa, a width of 296 nm, thickness of 220 nm, and length of 7.96 $\mathrm{\mu}$m. The total mass accounts for the additional width of the paddle and the air gaps which comprise the optical cavity; it is taken to have a value of 2.663 pg. Solving \autoref{suppEq:EBtheory} under these approximations yields predicted mode frequencies of 20.5 MHz, 56.6 MHz, 111 MHz, and 183 MHz for the first four modes in the resonance ladder. The fundamental torque mode at 3 MHz is not considered in this analysis, as Euler-Bernoulli theory considers only flexural bending modes. 

\subsection{Experimental Mode Frequencies} \label{suppSect:MechFrequencies}
\begin{figure*}[ht] 
	\centering
    \includegraphics[width=0.95\linewidth] {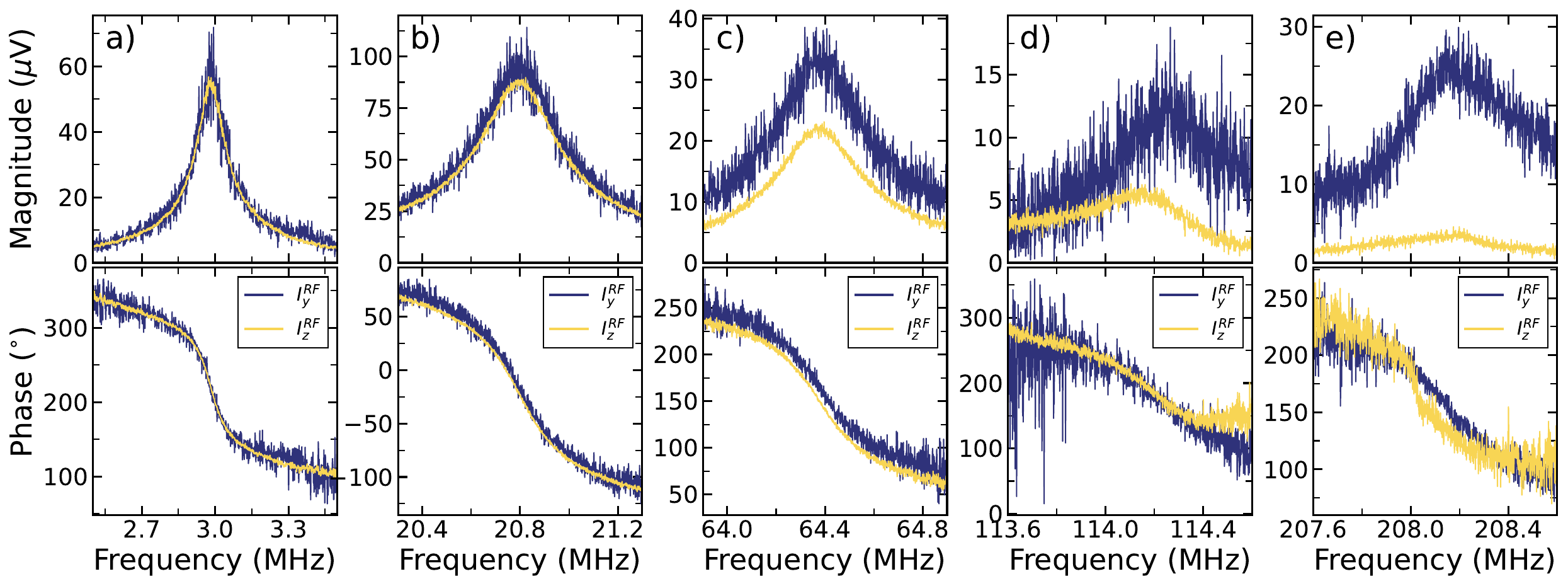}
	\caption{\textbf{$I_z^{\mathrm{RF}}$- and $I_y^{\mathrm{RF}}$-driven magnitude and phase.}
	Lock-in frequency sweeps of each mechanical mode after phasor subtraction of the background signal (as described in \autoref{suppSect:SignalProcessing}). Phase behaviour is included in the lower panels to illustrate the 180$^{\circ}$ phase shift expected across a resonance.  Direct comparison of magnitudes (top panels) is accomplished by determining the relative mechanical transduction from each coil via the 3 MHz $I_z^{\mathrm{RF}}$- to $I_y^{\mathrm{RF}}$-driven amplitude ratio (0.172). This same scaling factor is applied to each of the $I_z^{\mathrm{RF}}$-driven sweeps. The mechanical quality factors are measured to be (a) 36.1 $\pm$ 0.6, (b) 107 $\pm$ 2, (c) 301 $\pm$ 8, (d) 360 $\pm$ 20, and (e) 700 $\pm$ 40. }  
	\label{suppFig:FreqSweeps}
\end{figure*}

Mechanical resonances were determined through measuring signal as a function of frequency. Background subtracted frequency sweeps (subtraction described in \autoref{suppSect:SignalProcessing}) are shown in \autoref{suppFig:FreqSweeps} along with the corresponding phase for both drive coils. Resonances were measured on the decreasing hysteresis branch at a field of 14.89 kA/m. A signature of mechanical resonance is a 180$^{\circ}$ phase shift across the resonant peak. The central frequencies of each resonance correspond closely to mechanical frequencies calculated with COMSOL eigenfrequency analysis (3.9 MHz, 22.3 MHz, 65.2 MHz, 118.7 MHz, and 209.1 MHz).  The $I_z$-driven signal at 208 MHz is nearing the limit of detection of the experiment.

\section{Macrospin Simulation} \label{suppSect:Macrospin}

A macrospin model was developed to explore the frequency dependence of the mechanical torques predicted to arise from combined Einstein-de Haas and cross product magnetic torques.  Comparisons can be made between experimental data and the macrospin model when the sample is nearly saturated and only small closure domains prevent it from being uniformly magnetized.  In this limit we assume that the time-evolution of magnetization can be approximated as the time-evolution of a single giant spin with magnetic moment equal to $M_s$, where $M_s$ is the magnetization of the sample.  The macrospin model is a helpful, faster-executing complement to micromagnetic simulation packages which solve the Landau-Lifshitz-Gilbert (LLG) equation for a collection of interacting spins. 

The time-evolution of the driven macrospin is calculated with the LLG equation,
\begin{equation}
\frac{d\mathbf{m}}{dt} = -\gamma \mathbf{m} \times \mu_0 \mathbf{H}_{\mathrm{eff}} - \frac{\alpha}{M_s} \mathbf{m} \times \frac{d\mathbf{m}}{dt},
\label{suppEq:LLG}
\end{equation}
where in contrast to the standard micromagnetic simulation packages w The effective field, $\mathbf{H}_{\mathrm{eff}},$ can be represented as a function of the energy density functional, $w$, which contains the total energy of the system. The relevant contributions for a magnetic system  with magnetization vector $\mathbf{m} = \mathrm{M}_{\mathrm{s}}\hat{m}$, applied field $\mathbf{H}_{\mathrm{app}}$, and demagnetizing field $\mathbf{H}_{\mathrm{d}}$ are the anisotropy ($w_A = -\hat{m}\cdot \mathcal{K} \cdot \hat{m}$), demagnetization ($w_D = -\frac{1}{2}\mu_0\mathbf{m}\cdot \mathbf{H}_{\textrm{d}}$), and Zeeman ($w_Z = -\mu_0\mathbf{H}_{\textrm{app}}\cdot\mathbf{m}$) energy densities. The diagonal tensor $\mathcal{K}$ defines the easy anisotropy axis such that the easy axis has value $K_u$ for uniaxial anisotropy, with all other entries equal to zero. 

The solution to the LLG equation in spherical coordinates,
\begin{equation}
\begin{bmatrix}
\theta \\
\phi
\end{bmatrix} = \frac{-\gamma}{M(1+\alpha^2)} \begin{bmatrix}
\alpha & \frac{1}{\sin\theta}\\
\frac{-1}{\sin\theta} & \frac{\alpha}{\sin^2\theta}
\end{bmatrix} \begin{bmatrix}
\frac{\partial w}{\partial \theta}\\
\frac{\partial w}{\partial \phi}
\end{bmatrix} \label{suppEq:LLGsystemOfEquations}
\end{equation}
is expressed in terms of the total energy density functional, $w = w_A + w_D + w_Z$, the gyromagnetic ratio, $\gamma$, and the Gilbert damping constant, $\alpha$ for magnetization vector $\mathbf{m} = (\mathrm{M_s},\theta, \phi)$. Numerical integration of \autoref{suppEq:LLGsystemOfEquations} yields the time evolution of spherical angles $\theta$ and $\phi$ given initial magnetization conditions and simulation parameters including external DC and RF fields, material parameters, anisotropy, and sample shape \cite{GitHub_Macrospin}.  The LLG solution is further processed to compute reaction torques resulting from the precession of magnetic moment. These reaction torques can be directly inferred from the LLG equation by expressing \autoref{suppEq:LLG} as an equality between a driving torque ($\mathbf{\tau} = \mathbf{m} \times \mu_0\mathbf{H}_{\mathrm{ext}}$) and a net reaction torque through breakdown of the effective field into external ($\mathbf{H}_{\mathrm{ext}}$) and internal ($\mathbf{H}_{\mathrm{anis}}$) contributions. In this way the LLG equation may be expressed as
\begin{equation}
\mathbf{m} \times \mu_0 \mathbf{H}_{\mathrm{ext}} = -\frac{1}{\gamma}\frac{d\mathbf{m}}{dt} -\mathbf{m} \times \mu_0 \mathbf{H}_{\mathrm{anis}} - \frac{\alpha}{\gamma M_s} \mathbf{m} \times \frac{d\mathbf{m}}{dt}
\label{suppEq:LLG_rearrange}
\end{equation}
which indicates three contributions to the net reaction torque per unit volume,
\begin{equation}
\tau^{\mathrm{drive}} = \tau^{\mathrm{reaction}}= \tau^{\mathrm{ellip}} + \tau^{\mathrm{anis}} + \tau^{\mathrm{damp}}.
\label{suppEq:LLG_rxnTorques}
\end{equation}
The reaction torque component per unit volume $\tau^{\mathrm{ellip}} = -\frac{1}{\gamma}\frac{d\mathbf{m}}{dt}$ describes the precession of magnetization in an elliptical orbit. This torque is (similar to the EdH torque) frequency dependent and retains its elliptical orbit, never falling onto a single-axis path. Of the remaining magnetic torque components, the anisotropy torque describes the magnetic precession induced by internal anisotropy fields and the damping torque describes the decay of the magnetization vector towards its equilibrium direction. The net reaction torque per unit volume $\tau^{\mathrm{reaction}}$ is superimposed with the EdH torque per unit volume, $\tau^{\mathrm{EdH}} = -\frac{1}{\gamma '}\frac{d\mathbf{m}}{dt}$ which is not described by the LLG equation, to reveal the net magnetic torque.

\subsection{Lock-In Emulation} \label{suppSect:LockIn}
Direct comparison of sinusoidal simulation outputs resulting from AC external fields to measured torques demands the rms magnitudes and phase of the reaction torques. These parameters are determined from simulation outputs by filtering the solutions with the same techniques employed by lock-in amplifiers. The simulation output is mixed with a reference function, $V_r(t) = \sqrt{2}e^{i\omega_rt}$ before applying an nth order low-pass filter with transfer function
\begin{equation}
H_n(\omega) = \frac{1}{\left(1+i\frac{\omega}{\omega_{\mathrm{cutoff}}}\right)^n}
\end{equation}
to produce the root-mean-squared magnitude and phase. Simulations in this work are processed with a fourth order filter. For simulations performed with time-dependent frequency $\omega_r(t)$, the reference function is defined as $V_r(t) = \sqrt{2}e^{i\int_0^t \omega_r(t)dt}$. The same filtering techniques are applied to both macrospin and mumax simulation outputs.

\section{Local inductive probe measurements} \label{suppSect:SnifferCoil}
To test against the possibility of screening currents (in the silicon handle layer of the device or anywhere in the sample mounting fixture) introducing a frequency dependence to the relative amplitudes and phases of the $H_y^{\textrm RF}$ and $H_z^{\textrm RF}$ fields, a 1 mm diameter, two-turn pickup coil was used to detect the RF fields from the transmission line board, with and without the silicon chip in position on the board. The magnitude and phase of the inductive signal was measured with an SR844 RF lock-in amplifier, with the plane of the pickup coil parallel (perpendicular) to the surface of the board for coupling to $H_z^{\textrm RF}$ ($H_y^{\textrm RF}$).  Recorded magnitudes and phases are shown in \autoref{tab:SnifferCoilData} for the five mechanical mode frequencies of the experiment (with 200 MHz used in place of 208 MHz here, owing to the SR844 lock-in frequency limit). For both sets of coils, the difference in magnitude and phase measured by the pickup coil was within systematic uncertainty relating to repositioning of the pickup coil for addition and removal of the silicon chip. The constancy of field indicates the silicon chip is transparent to RF fields up to 200 MHz and suggests no screening currents are present which affect the magnetic field experienced at the sample. Phase comparison between the two coils yielded minimal difference. Discrepancies in the phase measurements can be attributed to required repositioning of the sniffer coil to record perpendicular fields from the two coils.  

The inductive probe is not viable for establishing absolute phase references between the RF drive fields and detected mechanical torques, owing to very different phase delays in the two signal pathways.  A magneto-optical probe, where the RF field reference signal propagates through a matched optical fiber and photoreceiver, will be needed. 

\begin{table}[ht]
    \centering
    \footnotesize
    \setlength{\tabcolsep}{1pt}
    {\renewcommand{\arraystretch}{1}%
    \begin{tabular}{|c|c|c|c|c|c|c|c|c|c|}
        \hline
        % \multirow{2}{*}{Frequency (MHz)}
        Frequency
        &\multicolumn{2}{|c|}{$I_z^{\mathrm{RF}}$ (no chip)}
        &
        \multicolumn{2}{|c|}{$I_z^{\mathrm{RF}}$ (with chip)} 
        & 
        \multicolumn{2}{|c|}{$I_y^{\mathrm{RF}}$ (no chip)} 
        & 
        \multicolumn{2}{|c|}{$I_y^{\mathrm{RF}}$ (with chip)} \\ 
        
        \cline{2-3}\cline{4-5}\cline{6-7}\cline{8-9}
         (MHz) 
         &Magnitude ($\mu$V) & Phase ($^{\circ}$)    &Magnitude ($\mu$V) & Phase ($^{\circ}$)    & Magnitude ($\mu$V) & Phase ($^{\circ}$) & Magnitude ($\mu$V) & Phase ($^{\circ}$)\\
        \hline
        3&   6.77&71.3  &6.865&71.4  &2.74&71.2  &2.74&71.2\\
        21&  33.03&8.98  &33.5&9.05  &13.23&9.6  &13.27&9.78\\
        64&  114.1&-148.1  &115.7&-147.9  &46.7&-145.1  &46.7&-144.99\\
        114& 165.4&39.3  &167.6&39.07  &68.38&46.2  &68.52&45.5\\
        200& 252.0&80.6  &255.3&80.5  &106.6&87.9  &107.0&87.35\\
        \hline
    \end{tabular}}
    \caption{Magnitude and phase data recorded for measurements of transmission line coils with and without a Si chip present. Small differences in magnitude and phase between the same drive coil with and without the Si chip in place are attributed to system noise and uncertainty resulting from repositioning the coil to add and remove the Si chip. }
    \label{tab:SnifferCoilData}
\end{table}

\section{Signal Processing} \label{suppSect:SignalProcessing}
\subsection{Phasor Subtractions} \label{suppSect:PhasorSubs}
\begin{figure*}[ht] 
	\centering
    \includegraphics[width=0.75\linewidth]{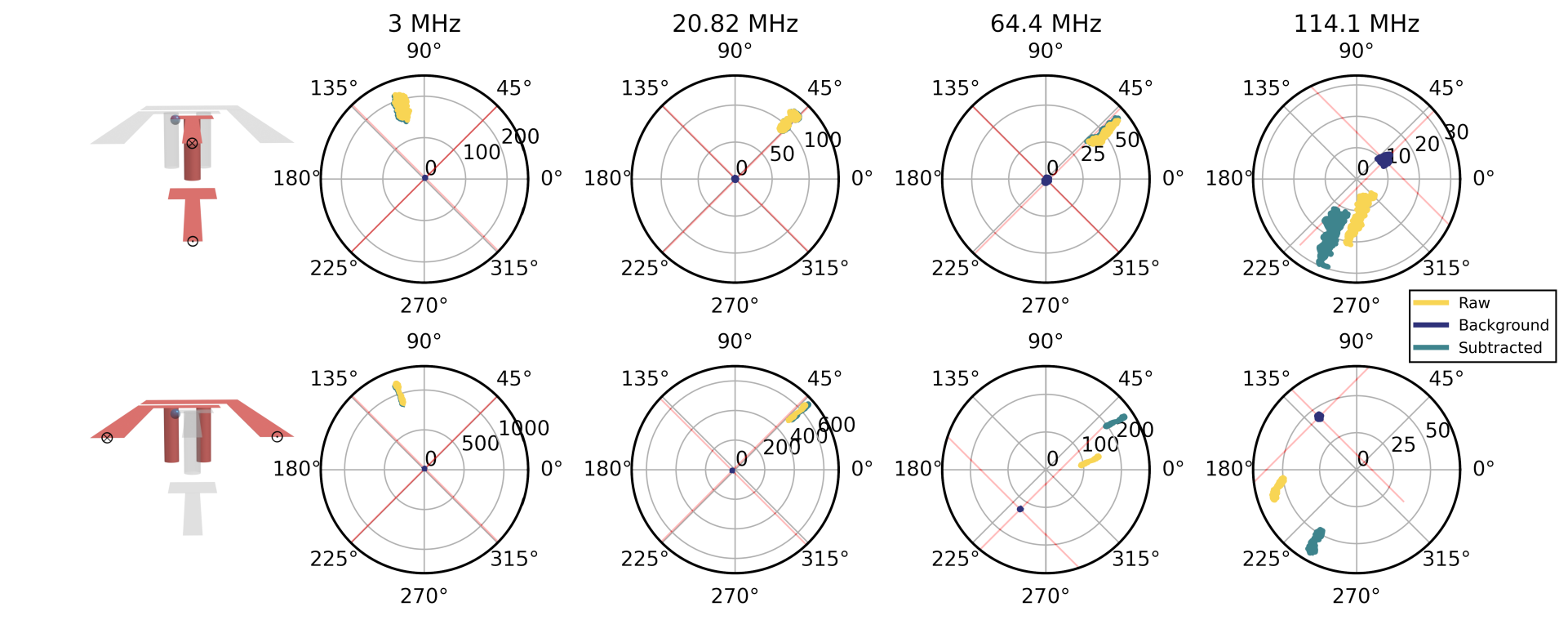}%{PhasorSubIllustration_polarAxes_v3.png}
    \caption{\textbf{Polar representation of raw, background, and subtracted signals.} For both drive field orientations at 3, 21, 64, and 114 MHz, average measured torques between 50 and 10 kA/m are illustrated on polar axes. The subtracted torques can be visualized as a shifting of the curves such that the background is centered on the origin. The red markers indicate the ``new" origin that is created during background subtraction. Low field hysteresis contains large phase and magnitude variations that shroud the procedure and are not shown.}
    \label{suppFig:PhasorSub}
\end{figure*}

Torque signals and background (tuned away from the optical cavity resonance) signals are each recorded as in-phase and quadrature components through lock-in amplification. As such, removal of background from net signal requires processing similar to vector subtraction in order to maintain accurate phase information. The in-phase and quadrature components of the background signal are subtracted from the same components of the net torque signal. Magnitude and phase are calculated using standard vector analysis. Polar coordinates provide a helpful visualization tool to describe this process.  The background subtraction can be imagined as a reorientation of the axes to align the origin with the background, as illustrated in \autoref{suppFig:PhasorSub}. For small backgrounds, this process has little effect.  Background signal increases with frequency due to RF noise, resulting in a greater impact of the background subtraction procedure on higher frequency measurements. As part of this subtraction, the phases of all curves are offset such that the initial phase (i.e. high field region) is zero. This is done to enable comparison between all phases for an apparatus in which the phase of the drive field at the sample is not known and thus the relative phase between the lock-in output and the measured signal is not an absolute measure of the phase in the system. 

\subsection{Normalization of Hysteresis} \label{suppSect:Normalization}
Direct comparison of torque for multiple mechanical mode orders requires normalization owing to decreased mechanical transduction for higher-order modes. Because conventional cross product torques are frequency-independent (\autoref{fig:macrospin}a of the main text) and the EdH contribution is negligible at high fields even for higher order mechanical modes, the $\mathrm{I_z^{\mathrm{RF}}}$-driven torques are scaled by their high-field magnitude (1104 $\pm$ 8 $\mu$V, 656 $\pm$ 3 $\mu$V, 248 $\pm$ 1 $\mu$V, and 62.1 $\pm$ 0.9 $\mu$V for 3, 21, 64 and 114 MHz respectively). Mechanical transduction of torque is frequency-dependent but drive-independent. The $\mathrm{I_y^{\mathrm{RF}}}$-driven torques can thus be scaled by the same factors to illustrate the relative magnitudes of the torques as a function of frequency.

\subsection{Micromagnetic Simulation of Hysteresis} \label{suppSect:MumaxHysteresis}

\begin{figure*}
    \centering
    \includegraphics[width=0.8\linewidth]{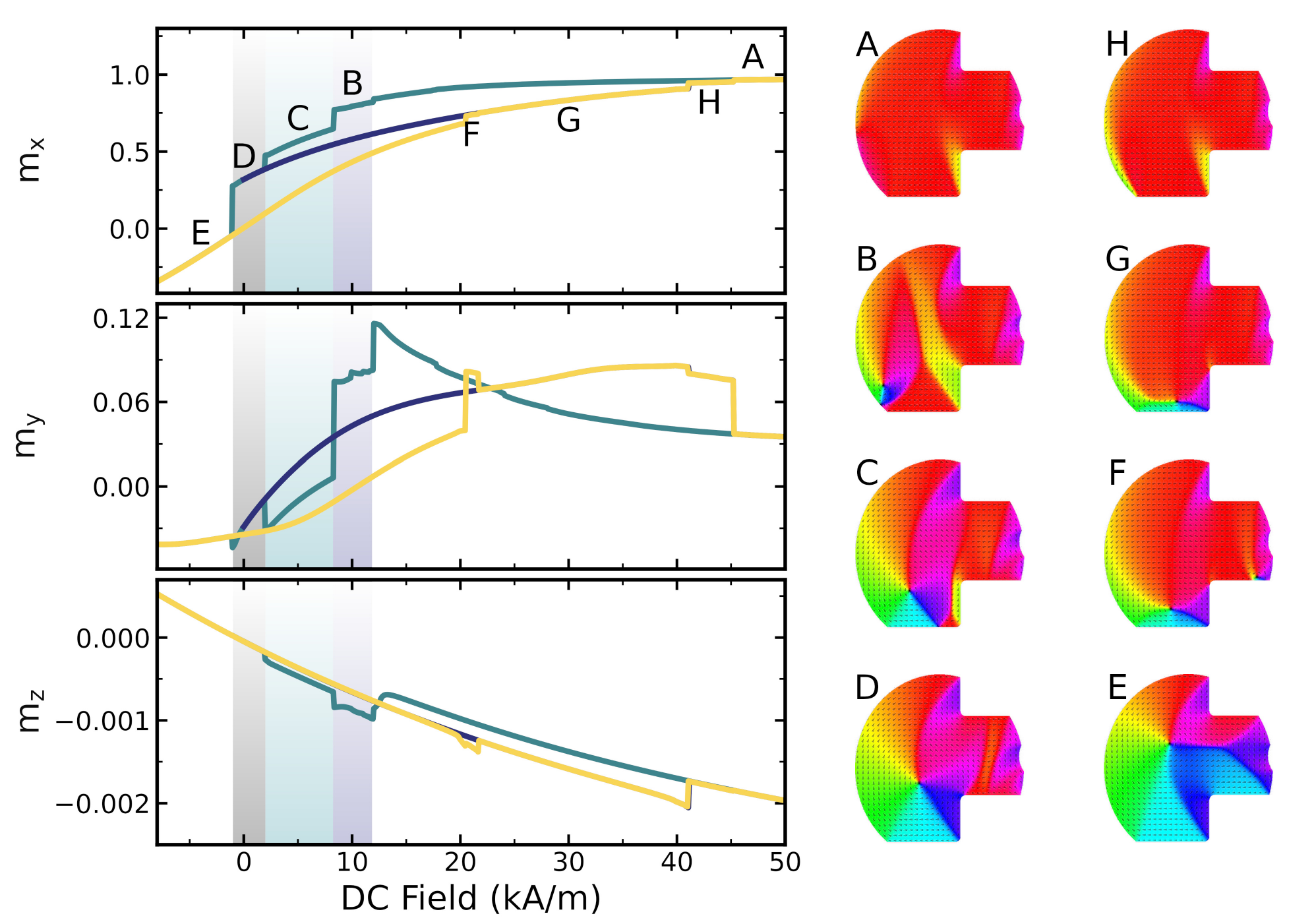}%{simulatedHysteresis_withSpinTextures-v5.png}
    \caption{\textbf{Simulated Hysteresis for permalloy film.} 
    The three components of magnetization are shown for sweep direction reversal at 0 (purple) and -7.96 (yellow) kA/m for a simulated hysteresis loop beginning with downward sweep (green) at 50 kA/m. Spin textures throughout the loops are shown on the right. The fields they correspond to are indicated on the $m_x$ axis. Shading on the axes matches the shading included in \autoref{fig:hysteresis} to indicate distinct spin textures in measured hysteresis.}
    \label{suppFig:mumaxDChyst}
\end{figure*}

To understand the behaviour underlying measured torques, we rely on micromagnetic simulations to predict the outcome of DC hysteresis measurements. Such simulations have been successful in the past to describe the behaviours of conventional cross product torques, as well as some finite characteristics of EdH torques through dithering the drive field to mimic an alternating field (e.g. Ref \cite{Mori2020}). Micromagnetic simulations performed with Mumax$^3$ \cite{Vansteenkiste2014} are initialized as a saturated state in high field and allowed to relax to the minimum energy configuration. The DC field applied in the simulations reproduces the measured three-axis field produced by the biasing magnet at the sample location, with small $y$ and $z$ contributions defined by a power law for an $x$-field (in kA/m):
\begin{equation*}
    \mathbf{H} = \begin{cases}
    H_x\\
    4.329\times10^{-6}H_x^3 - 4.25\times10^{-4}H_x^2 + 4.125\times10^{-2}H_x - 0.1\\
    -2.181\times10^{-6}H_x^3 + 4.253\times10^{-4}H_x^2 - 4.507\times10^{-2}H_x + 8.982\times10^{-2}
    \end{cases}
\end{equation*}
The $H_x$ field value is stepped discretely and the energy is minimized for each step. The resulting magnetization vector is shown in \autoref{suppFig:mumaxDChyst}, along with the magnetic spin texture at key points along the hysteresis loop for field reversal occurring at 0 kA/m (purple curve) and -7.96 kA/m (yellow curve). The $m_x$ component shows directly comparable behaviour to measured magnetic torques (\autoref{fig:hysteresis}a-b in the main text), as expected for conventional cross product torques ($\mathbf{\tau} = \mu_0(\mathbf{m} \times \mathbf{H})$). Steps in $m_x$ appear in the same field range as observed in measured torque data, highlighted by backgrounds of the same colour in the main text. That the transitions do not occur at the same frequencies in simulation and experiment is expected, due to missing information (e.g. temperature, edge roughness, surface defects, etc.) in the simulation. 

The high-field range of the $m_y$ component reveals some similarity to the $I_y^{\mathrm{RF}}$-driven torques in \autoref{fig:hysteresis}c-d; pointing to increased $m_y$ with decreasing field. This behaviour is amplified in EdH $y$-torques, described by the time rate of change of $m_y$, for which greater excursion is permitted as $m_x$ is reduced with decreasing $H_x^{\mathrm{DC}}$ fields.

\subsection{Hysteresis at 3 MHz}
\label{suppSect:3MHzHysteresis}

\begin{figure}
    \centering
    \includegraphics[width=0.6\linewidth]{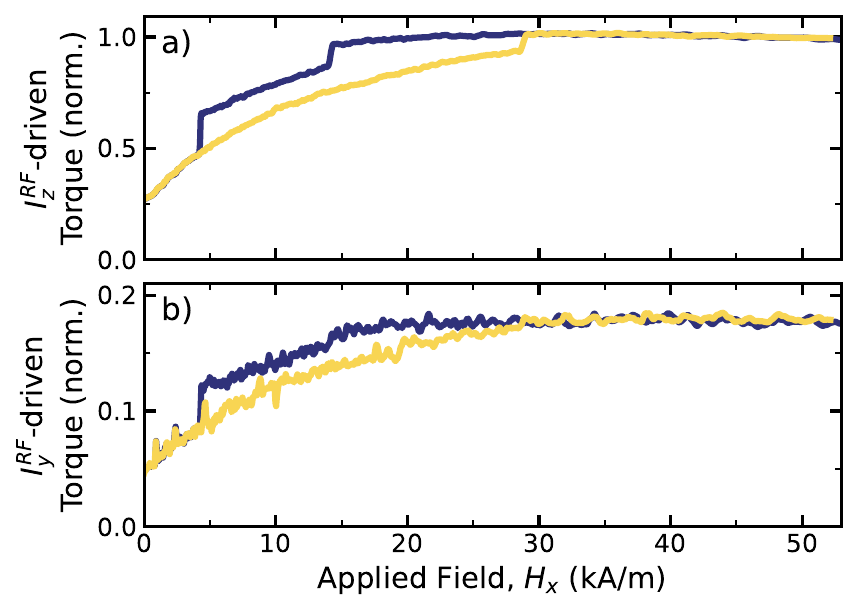}
    \caption{\textbf{3 MHz hysteresis for $I_y^{\mathrm{RF}}$ and $I_z^{\mathrm{RF}}$ drives.} The 3 MHz hysteresis curves for the $I_z^{\mathrm{RF}}$ (a) and $I_y^{\mathrm{RF}}$-drive (b) of \autoref{fig:hysteresis}, following the same normalization. Decreasing field is shown in purple, while increasing field is shown in yellow for both drives.  }
    \label{suppFig:3MHzHyst}
\end{figure}

At 3 MHz, the EdH contribution to torque is expected to be minimal in comparison to the cross-product contribution. Thus, the $I_y^{\mathrm{RF}}$-driven hysteresis curve follows the same behaviour as the $I_z^{\mathrm{RF}}$-driven curve, as illustrated in \autoref{suppFig:3MHzHyst}. While the $I_y^{\mathrm{RF}}$-driven curve has worse signal-to-noise owing to lower $H_z^{\mathrm{RF}}$ amplitude from the $I_y^{\mathrm{RF}}$ coil, the overall shape of the curves are nearly identical. This similarity allows identification of relative drive amplitudes of the coils (\autoref{suppSect:RFAdmixture}) and highlights the frequency-dependence of the EdH component from $H_y^{\mathrm{RF}}$ in \autoref{fig:hysteresis}. 

\subsection{Individual Hysteresis Measurements} \label{suppSect:IndividualHysteresis}

\begin{figure*}
    \centering
    \includegraphics[width=1\linewidth]{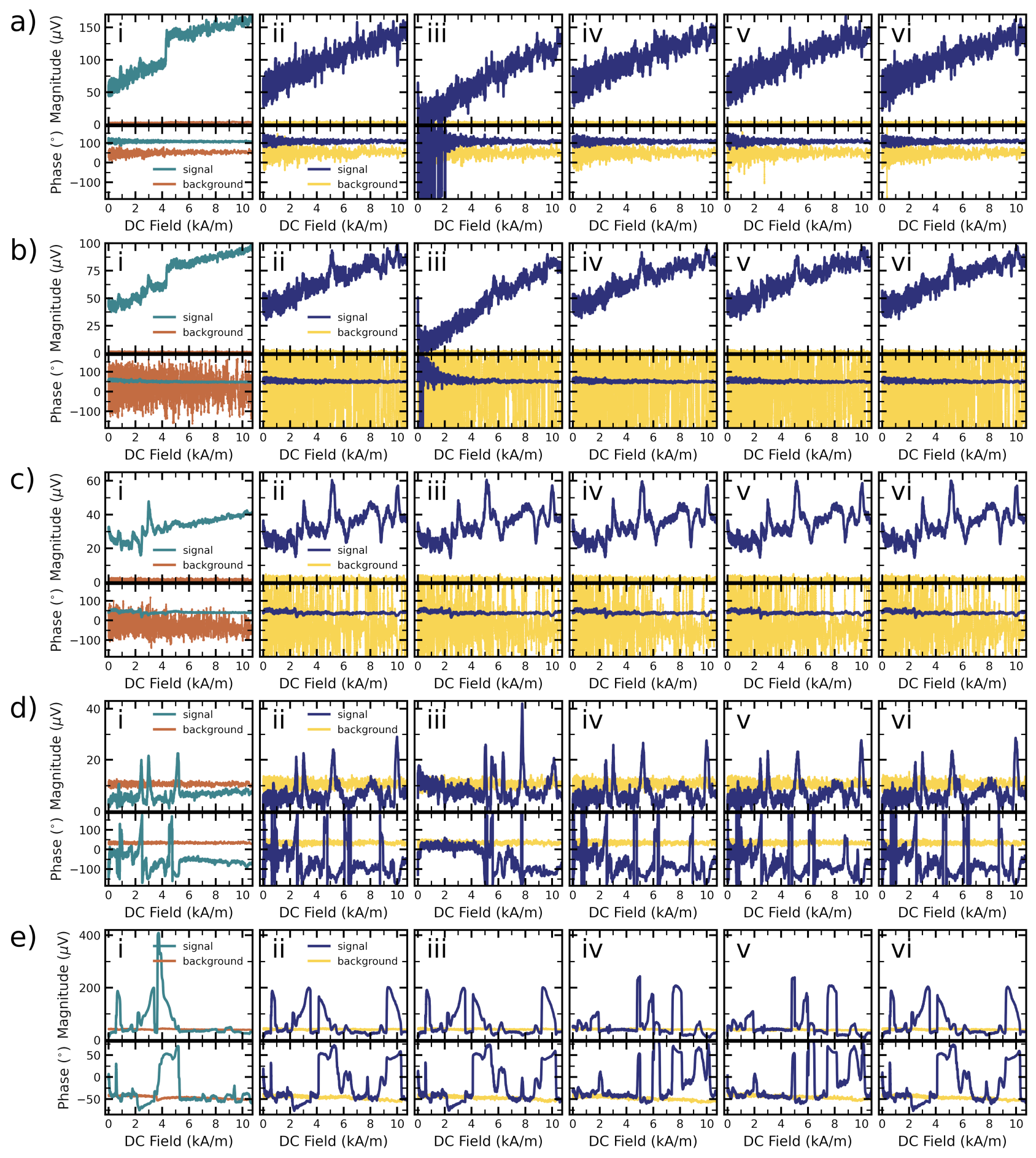}%{IndivSweeps_wHtLavg_v5_allFreqs_subplotversion_10x12FigSize.png}
    \caption{\textbf{Individually collected $I_y^{\mathrm{RF}}$-driven hysteresis.} Net signal and phase for a sequence of five consecutive hysteresis loops are shown for DC fields less than 10 kA/m along with background signal and phase for drive frequencies of (a) 3 MHz (b) 21 MHz (c) 64 MHz (d) 114 MHz (e) 208 MHz. The average high-to-low (HtL) field components are presented in  column i of each set.  Individual sweeps from low-to-high (LtH) fields are presented in the columns ii - vi of each set. Jumps in magnitude and phase relative to the HtL low-field torque indicate the presence of an additional spin texture ((a) iii, (b) iii, (d) iii, (e) iv, v).}
    \label{suppFig:Individual-Sweeps}
\end{figure*}

Hysteresis data seen in \autoref{fig:hysteresis} and \autoref{fig:magnres} of the main text were collected as a sequence of five consecutive loops. Comparison of the constituent loops of each sequence reveals an additional spin texture transition at very low field for select sweeps. Hysteresis sequences for an $I_y^{\mathrm{RF}}$ drive current are presented in \autoref{suppFig:Individual-Sweeps} for 3, 21, 64, 114, and 208 MHz driving frequencies. Of the 25 total loops, five show an additional spin texture characterized by a jump in magnitude and phase near the reversal of the DC field sweep direction. The transition is seen in the second loops for the 3 MHz, 21 MHz, and 114 MHz drive frequencies, and the third and fourth loops for the 208 MHz drive frequency. None of the $I_z^{\mathrm{RF}}$-driven hysteresis loops indicate the presence of this additional spin texture. The jump in measured phase as well as magnitude in the measured torques can be attributed to a sign change in the magnetization, as seen in the micromagnetic simulations with negative field reversal (\autoref{suppFig:mumaxDChyst}). 

\begin{figure*}[ht]
    \centering
    \includegraphics[width=1\linewidth]{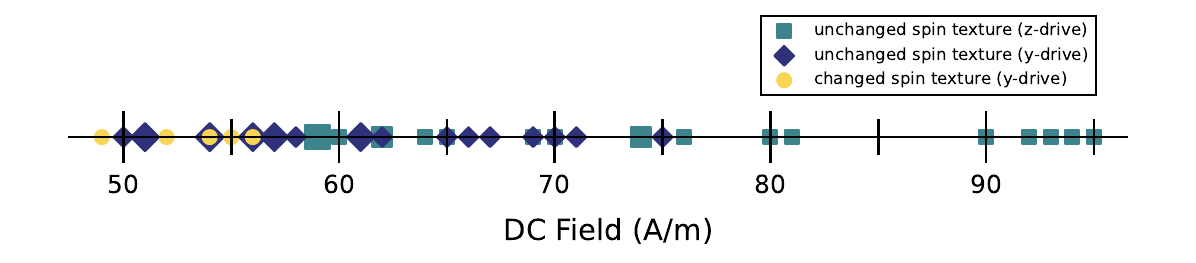}%{spinTexture-lowFieldHyst-fieldDependence_wZfields_v4.png}
    \caption{\textbf{Field-dependence of appearance of additional spin texture at low field.} The minimum fields observed for each individual hysteresis loop presented in \autoref{suppFig:Individual-Sweeps} are presented. Instances that indicate a zero-field spin texture transition are indicated by yellow circles. Instances with no low-field spin texture transition are marked by purple diamonds (y-drive) and green squares (z-drive). The changed spin texture shows a clear dependence to occur at the lowest values of DC field strength. Unchanged spin textures appearing at similar fields indicates that the spin texture is induced simultaneously by field strength and stochastic temperature fluctuations.}
    \label{suppFig:spinTexture-fieldDep}
\end{figure*}

Micromagnetic simulations predict a reversal around - 1 kA/m. The experimental transition evidently occurs closer to zero field and is likely driven primarily by stochastic temperature fluctuations when the field is sufficiently low. The recorded fields have very poor signal to noise near this transition due to their small amplitudes; however recorded values at the reversal point demonstrate clear variability in the field strength at the sample. Fluctuations in field minima (calculated as the mean recorded field around the point of reversal, with signal noise characterized as 30 A/m) can be attributed to small variations in the position of the permanent magnet owing to variability of the attached stepper motor. The transition fields for each loop are presented in \autoref{suppFig:spinTexture-fieldDep} for both drive coils. Fields corresponding to a zero-field transition are marked in yellow, illustrating a clear tendency to transition at lower field strengths. 

Comparing the field span encompassing the transition fields for this zero-field transition (7 A/m) with the corresponding spans for the 14.2 kA/m transition (150 A/m span) and 4.4 kA/m transition (110 A/m) in the high-to-low field hysteresis branch, it is clear that we are reaching the onset of fields which result in a transition at the limit of our permanent magnet's range of motion. The fields for the 14.2 and 4.4 kA/m transitions exhibit a general frequency-dependence. This indicates that the RF fields affect the spin texture in a similar manner to temperature, effectively heating the spins to produce the systematically increasing (for high-to-low field sweep) average transition field value with higher drive frequency. 

Despite small variation in transition fields, the sweeps without the zero-field spin texture transition show remarkable repeatability in the recorded features. Taking advantage of this to improve signal-to-noise, the average of the four sweeps with a consistent spin texture and the correspondingly averaged background is utilized in the phasor subtraction procedure described in \autoref{suppSect:PhasorSubs}. This averaged and subtracted data is presented in the main text Figure \autoref{fig:hysteresis}.  The same number of loops are averaged for the $I_z^{\mathrm{RF}}$-driven signals.  The 208 MHz hysteresis in \autoref{fig:magnres} contains the averaged signals from the first, second, and fifth loops presented in \autoref{suppFig:Individual-Sweeps}e. 

\subsection{Low-Field Hysteresis} \label{suppSect:lowFieldHysteresis}
\begin{figure}[ht]
    \centering
    \includegraphics[width=0.85\linewidth]{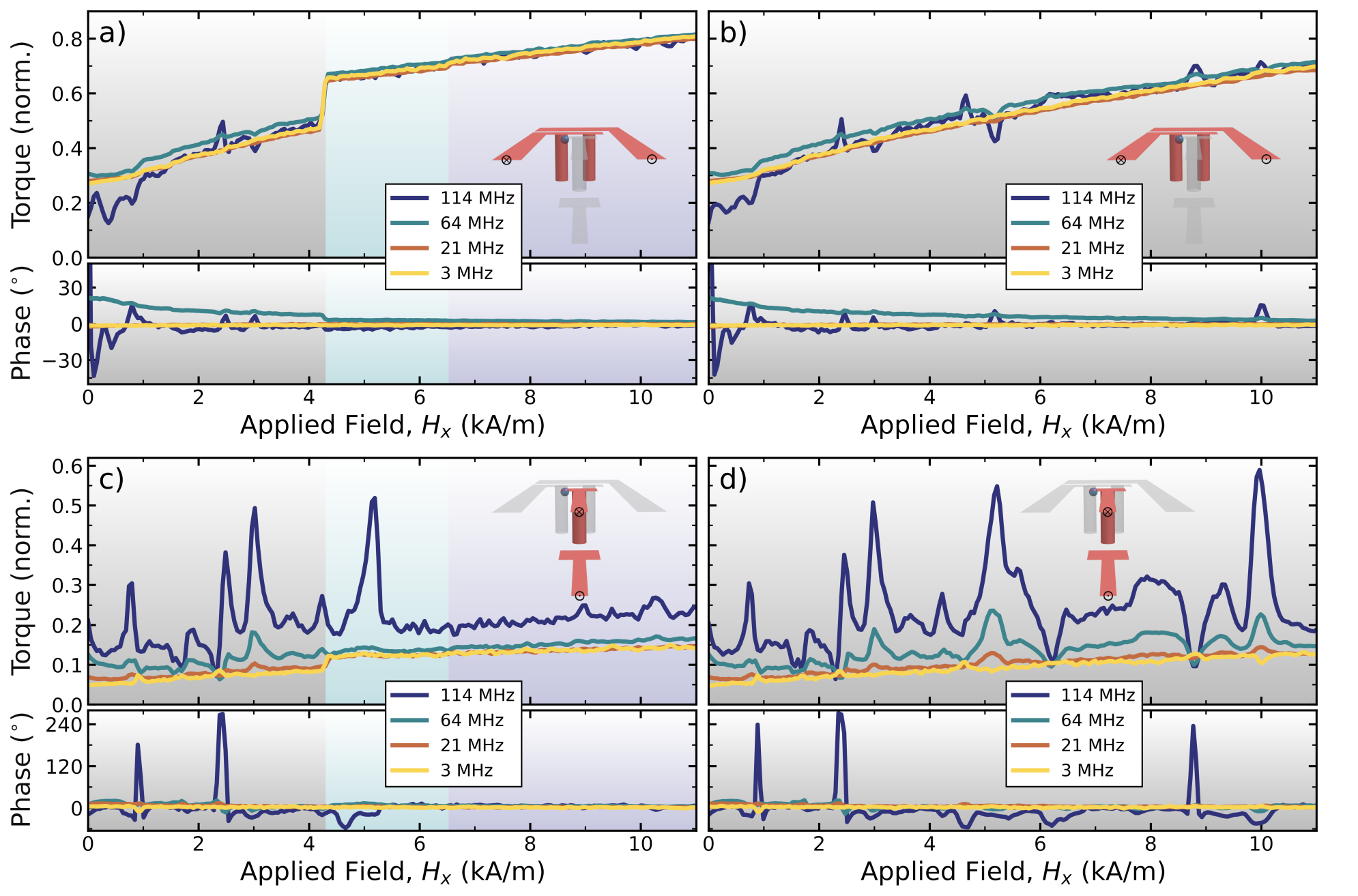}%{Hysteresis-lowFieldLimit-v4.png}
    \caption{Low-field limit of data presented in main text \autoref{fig:hysteresis} to highlight spikes in torque magnitude and phase.}
    \label{suppFig:LowFieldHysteresis}
\end{figure}

The averaged hysteresis data containing no zero-field transition (four scans) are illustrated in \autoref{suppFig:LowFieldHysteresis}. The subtle transition between a state with two vortices near the edge and a single vortex is evidenced by a slight decrease in torque amplitude at the transition point. In the single vortex spin texture, the $I_y^{\mathrm{RF}}$-driven torques exhibit large spikes in amplitude which are associated with spikes in phase. With the exception of the 64 MHz $I_z^{\mathrm{RF}}$-driven signal, which has a significantly larger background than any other signal (illustrated in \autoref{suppFig:PhasorSub}), the phase is mostly flat throughout the measurements, with the exception of the Barkhausen spikes. The large background in the $I_z^{\mathrm{RF}}$-driven 64 MHz signal leads to imperfect subtraction owing to increased signal present in the background measurement, producing the remnant of a transition in the phase of the subtracted signals. The large background in this measurement is attributed to RF interference during data collection. Such a signal does not appear in additional measurements of 64 MHz hysteresis.

\subsection{Admixture of RF Fields} \label{suppSect:RFAdmixture}
The position of the sensor relative to the coils produces admixture of $H_z^{\mathrm{RF}}$ from the $\mathrm{I_y^{\mathrm{RF}}}$ coil and vice versa, illustrated in \autoref{fig:fieldGeometry} of the main text. This admixture is advantageous for comparison of relative torque magnitudes between cross product and EdH torques. Using measured torque signals from each drive coil, we are able to determine the relative field amplitudes $H_y^{\mathrm{RF}}$ and $H_z^{\mathrm{RF}}$ produced by each coil.

At 3 MHz, the EdH torque is expected to be negligible relative to the cross product torque (\autoref{fig:macrospin}a) and as such, we consider the measured torque from the $I_y^{\mathrm{RF}}$ coil to be purely due to the $H_z^{\mathrm{RF}}$ field component. Indeed, when appropriately scaled, the 3 MHz torque signals from the $I_y^{\mathrm{RF}}$ and $I_z^{\mathrm{RF}}$ coils fall directly atop one another. The relative $H_z^{\mathrm{RF}}$ component from the $I_y^{\mathrm{RF}}$ coil is calculated to be 17.7 $\pm$ 0.4\% the strength of the $H_z^{\mathrm{RF}}$ component from the $I_z^{\mathrm{RF}}$ coil. 

The relative $H_y^{\mathrm{RF}}$ strengths are determined based on the small EdH behaviour observed in the 114 MHz $I_z^{\mathrm{RF}}$-driven torque signal. As discussed in the main text, Barkhausen jumps are clearly visible at low bias field strengths. An additional indicator of EdH torque is a slightly decreased magnitude near the first transition in the sweep down. Phasor subtraction of the scaled 3 MHz $I_z^{\mathrm{RF}}$-driven (purely cross product) torque from the 114 MHz $I_z^{\mathrm{RF}}$-driven torque reveals a torque curve with a clear signature of the expected ``ski jump" approaching the transition. The amplitude at the apex of the jump (around 14.5 kA/m) is compared with the 114 MHz $I_y^{\mathrm{RF}}$-driven torque (also phasor subtracted by the 3 MHz $I_z^{\mathrm{RF}}$-driven signal) to determine a $H_y^{\mathrm{RF}}$ ratio of 19 $\pm$ 9 between the $I_z^{RF}$ and $I_y^{RF}$ coils.

The data is not conducive to determination of relative field strengths for a single drive coil. To determine these ratios for use in simulations, finite-element simulation of the coils is performed and the field at the sample location is used to estimate the relevant ratio. Experimental field ratios between the two coils are compared to simulated field ratios at the sample location to verify accuracy of the simulations.

\section{$\mathrm{H_z^{\mathrm{RF}}}$-driven 208 MHz } \label{suppSect:208MHzHzRF}
\begin{figure*}[ht]
    \centering
    \includegraphics[width=0.7\linewidth]{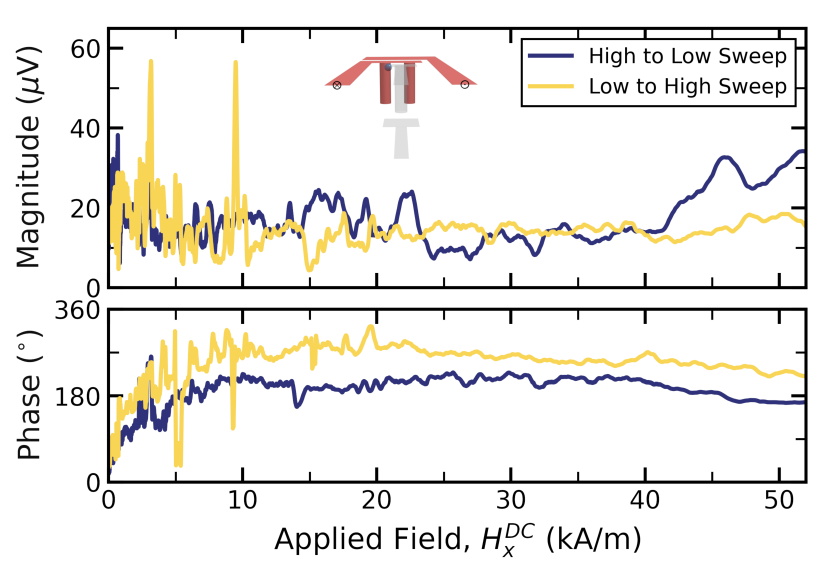}%{Supplementary Figures/208MHz_zdriven_0611_1stDsweep_5thBGsweep.png}
    \caption{\textbf{Hysteresis of z-driven 208 MHz torque.} Background-subtracted 208 MHz torque for a single hysteresis loop. Small mechanical transduction of higher order modes leads to small measured cross product torque amplitude. High background signal combines to produce poor signal to noise in comparison to the gyro-amplified $H_y^{\mathrm{RF}}$-driven signal. }
    \label{suppFig:zDriven208Hyst}
\end{figure*}

The cross product torque dominates over the EdH torque for all mechanical frequencies presented in \autoref{fig:hysteresis} in the main text. The macrospin model predicts this will continue past 208 MHz (\autoref{fig:macrospin}a main text). Empirically, mechanical transduction decreases approximately as $e^{-0.025f}$ with increasing frequency. This trend predicts the 208 MHz signal should not be visible above the noise floor, particularly when the RF noise increases with frequency, as observed in \autoref{suppFig:Individual-Sweeps}. The gyrotropic resonance near 208 MHz amplifies $H_y^{\mathrm{RF}}$-driven signals to vastly above the noise floor, but $H_z^{\mathrm{RF}}$-components do not couple to the gyrotropic mode.  As such, the cross product component of torque at 208 MHz remains buried in RF noise. A single $I_z^{\mathrm{RF}}$-driven 208 MHz torque measurement is shown in \autoref{suppFig:zDriven208Hyst}, where the only recognizable features in comparison with other hysteresis measurements of this device are the spikes present in the vortex configuration. These spikes are the result of $H_y$ admixture from the $I_z^{\mathrm{RF}}$ coil. All other features are indistinguishable from noise. 
%\mf{is the suggestion from the data, that there may be a high pass roll-off of the detected cross product mechanical torque amplitude (relative to the EdH contribution) also required here to make the argument self-consistent with the global analysis?  If the $I_y^{\mathrm{RF}}$-driven signal at high bias fields (ski slope) is still well above the 208 MHz noise floor then, all else being equal, so too would the $I_z^{\mathrm{RF}}$-driven torque?}

\section{Pinned Micromagnetic Simulations} \label{suppSect:PinnedSimulations}

The influence of a pinning potential on EdH torque amplitude near the gyrotropic frequency is investigated by modifying the saturation magnetization of small cylindrical regions along the path of the vortex core. The core trajectory is determined through DC hysteresis simulations in which an external field is applied along the $\hat{x}$-direction with discrete magnitude steps of 79.6 A/m (1 G). At each field magnitude, the system is allowed to relax to its equilibrium state. Pinning sites are introduced as 20 nm diameter cylindrical regions with saturation magnetization of 613.6 kA/m (80\% of $M_s$ for permalloy) along the path travelled by the vortex core. The path of the vortex core is determined by tracking the region corresponding to high out-of-plane magnetization. In the vortex configuration, spins are primarily oriented in the plane of a thin film sample, with out-of-plane magnetization only appearing at the vortex core. Thus, only the vortex core will produce significant out-of-plane magnetization, providing a signature to track.  With the pinning sites applied, DC hysteresis is repeated. The resulting spin textures from the second simulation are used in subsequent time-dependent simulations, wherein a $\hat{y}$-directed  10 A/m RF field is applied for 30 periods at each discrete DC field magnitude. 

Each discrete field step produces transients in the magnetization and torque evolution. These transients are purely the result of the stepped field; a smoothly varying field produces no such features. Thus, after applying the lock-in filtering techniques described in Section S2, the early-time solutions are neglected. Torque is calculated as the mean rms values of the late-time solutions (final 5000 data points). 

Gyrotropic orbits near the pinning sites are determined through ringdown similations performed similarly to those described in the main text. A 1 ns 0.398 A/m pulse is applied to the pre-calculated spin texture and the simulation is run for 25 ns. The position of the core is determined as described above to show the variation of orbital trajectory. Gyrotropic frequencies are determined through FFT transform of the magnetization. Unpinned simulations show roughly linear field-dependence of the gyrotropic frequency, whereas the pinned simulations reveal significant effect as the core interacts with the pinning site. 

\subsection{Simulations of Double Pinning} \label{suppSect:DoublePinnedSimulations}

\begin{figure*}[ht] 
	\centering
    \includegraphics[width=0.7\linewidth]{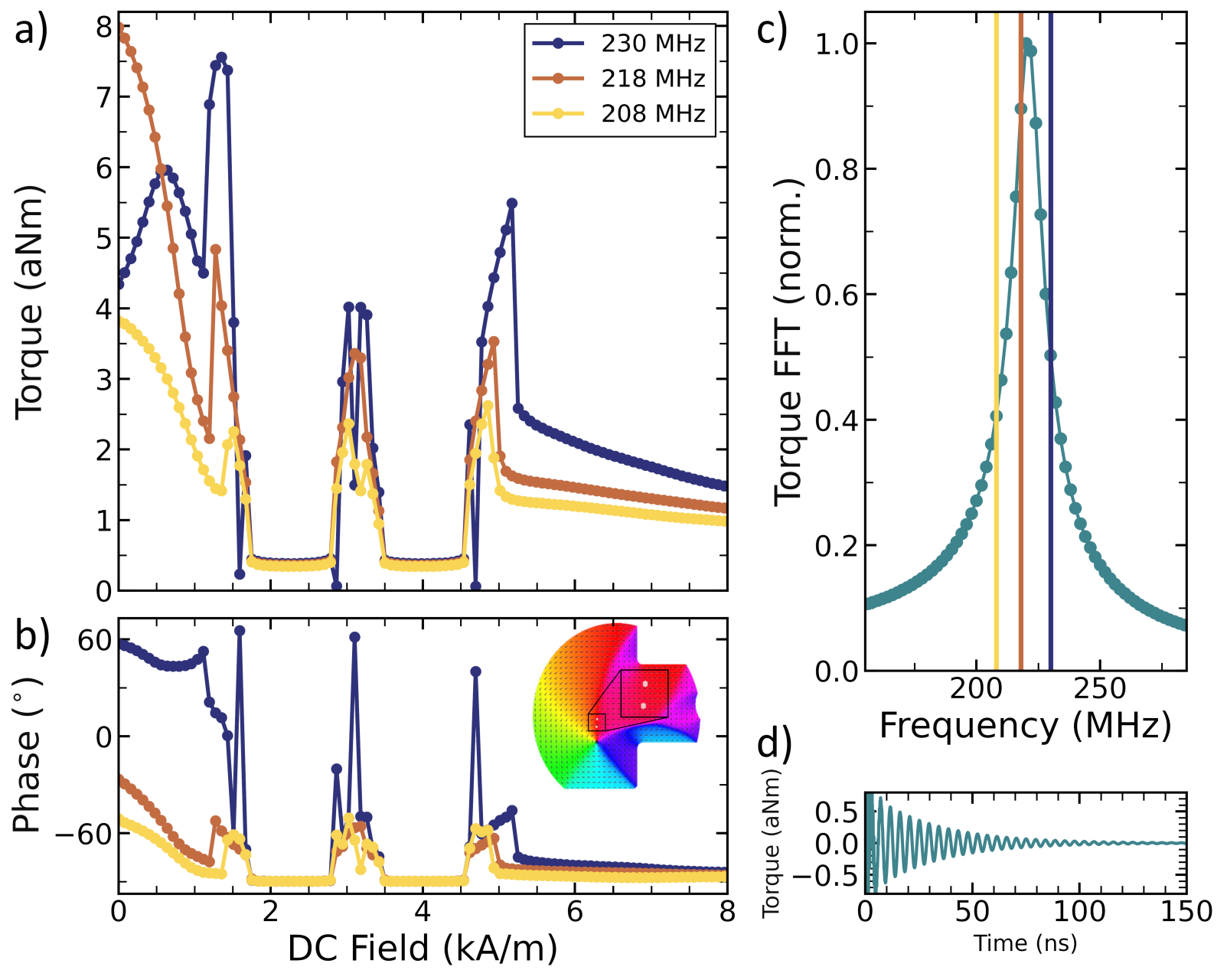}%{DoublePinning-MumaxSim-wRingdown_wPyShroomInset-v3.png}
    \caption{\textbf{Pinned Simulation.} Simulated torque magnitude (a) and phase (b) for a permalloy film in a vortex spin configuration. Two circular defects (20 nm radius) are applied along the vortex trajectory. Torques are calculated for three driving frequencies (208, 218, 230 MHz) to illustrate the effect of proximity to gyrotropic frequency of the film and the location of the two defects. The expanded region containing the defects is a 568.0x525.8 nm$^{\mathrm{2}}$ box. (c) The Fourier transform of a ringdown simulation performed at zero external DC field. The three frequencies used in (a) and (b) are indicated by lines of the same colours. (d) The time-evolution of the torque from the ringdown simulation. }
    \label{suppFig:Pinning-Simulation}
\end{figure*}

\begin{figure}[ht]
    \centering
    \includegraphics[width=\linewidth]{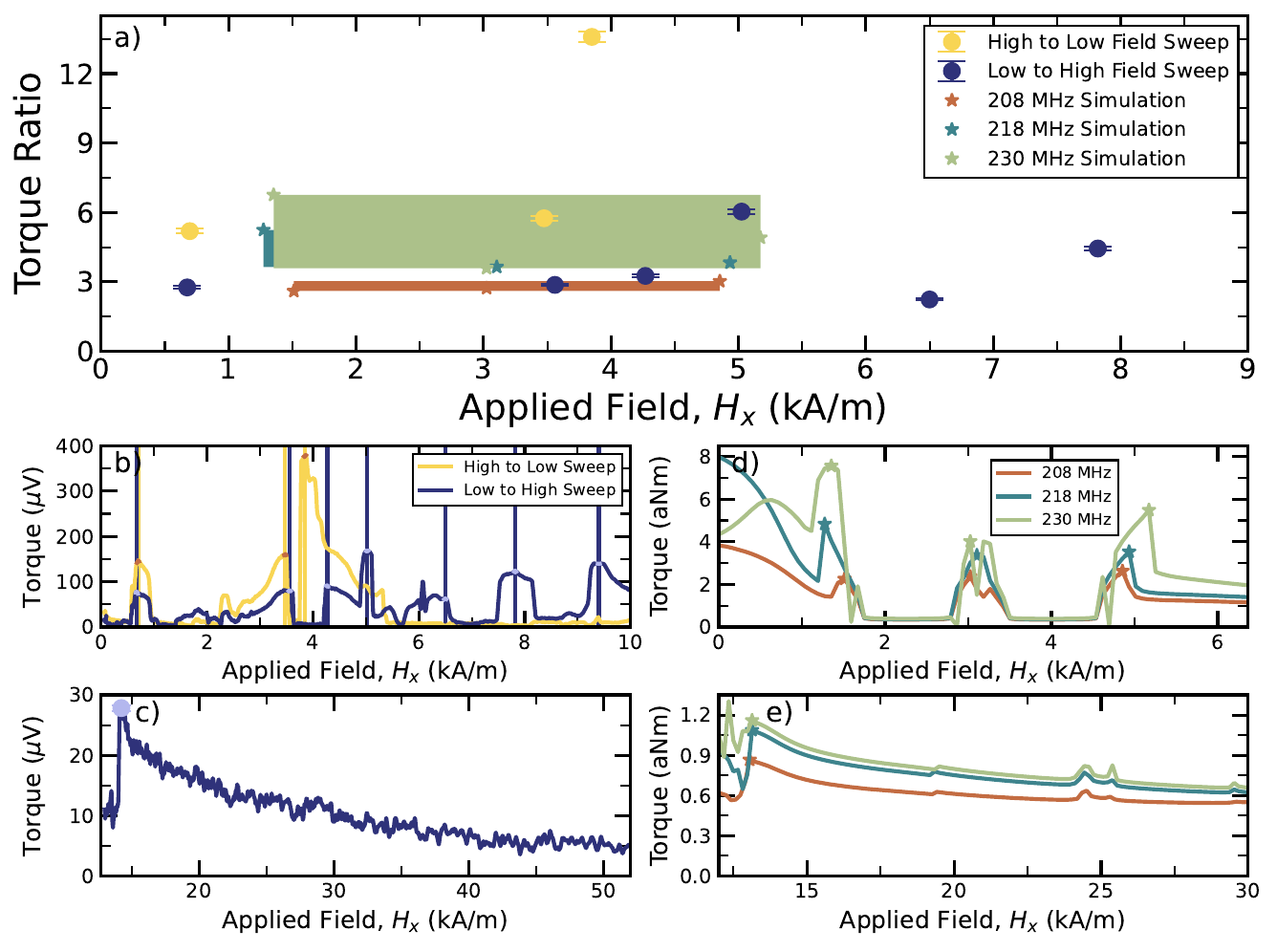}%{skiSlope_spike_Comparison-v4.png}
    \caption{(a) The ratio of torque magnitude of the low-field spikes (b,c) to the ``ski jump" feature of the magnetically saturated state (d,e) is compared between experiment (b,d) and simulation (c,e). The spikes in measured torque are determined as the mean within a narrow field range encompassing the peak of the spike. The field band for each spike is indicated in (b) by a shaded bar and discoloured data. The high-field maximum is calculated similarly. Simulations were performed in discrete field steps. Maxima are identified as single points for each simulation indicated by an x in (c0 and (e). }
    \label{suppFig:SkiSlopeSpikeRatio}
\end{figure}

Simulated torques for two pinning sites separated by 41 nm are shown in \autoref{suppFig:Pinning-Simulation} for three frequencies near the zero-field gyrotropic frequency. The torques exhibit large spikes in magnitude and phase when the vortex core is influenced by the pinning site but not fully captured. Inside the pinning site, the magnitude is negligible and the phase is fixed at -90$^{\circ}$. The behaviour of these torques is qualitatively similar to the observed behaviour of measured torque at 208 MHz (\autoref{fig:magnres} in main text). The distinct frequencies at which the simulations are performed are representative of field-dependent gyrotropic frequencies. \autoref{suppFig:Pinning-Simulation}c indicates 218 MHz is closest to the zero-field gyrotropic frequency, while 230 MHz corresponds to a higher-field gyrotropic mode. The field at which 230 MHz is the gyrotropic frequency is closest to the applied field when the core interacts with the first pinning site, amplifying the effect of pinning on the resultant torque. This correspondence is indicative of gyrotropic resonance producing the large low-field spikes observed only in 208 MHz hysteresis measurements. This correspondence is further investigated in \autoref{suppFig:SkiSlopeSpikeRatio} through comparison of the torque magnitude ratio of the spikes to the saturated maximum torque (``ski jump" feature) directly prior to the transition into a quasi-vortex spin texture. 

The higher DC fields present throughout the saturated regime produce gyrotropic resonances at higher frequencies. As such, the 208 MHz driving field does not co-resonate to greatly amplify torques in this range. To verify that the simulations are describing the measured torque behaviour, additional Mumax$^3$ simulations in the quasi-uniform regime were performed. These simulation were run with a smoothly varying field starting at 50 kA/m, ending just below the transition. The resulting amplitudes at the transition were compared to the amplitudes of the torque spikes resulting from the pinning simulations. The same amplitudes were compared for measured torques. As seen in \autoref{suppFig:SkiSlopeSpikeRatio}a, the ratio of torques between simulation and experiment yield good agreement, with the exception of the extremely large peak at 3.6 kA/m in the downward field sweep. This is indicative of a coincident gyrotropic mode with frequency 208 MHz at 3.6 kA/m.

\section{TMRS for 208 MHz}
\label{suppSect:TMRS}

\begin{figure}
    \centering
    \includegraphics[width=0.8\linewidth]{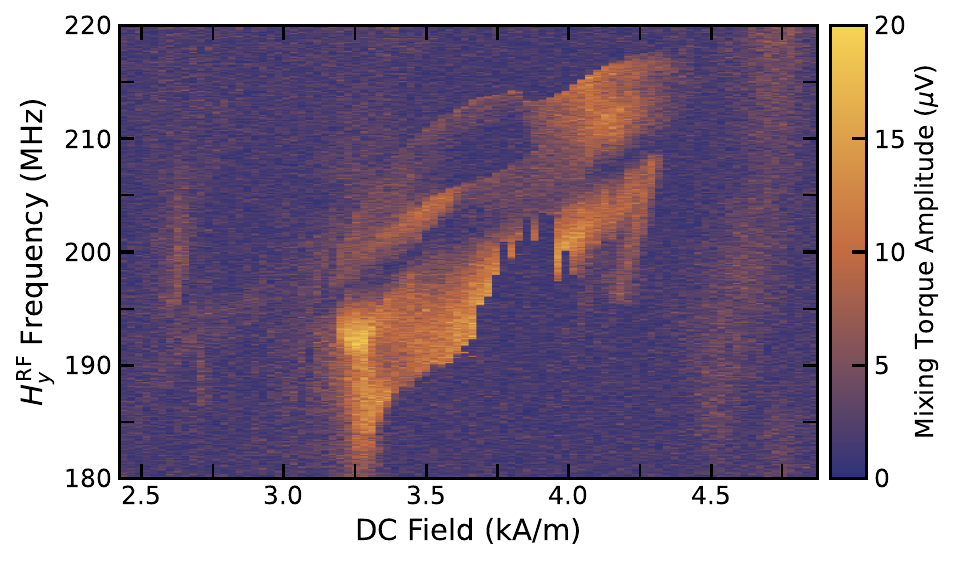}%{TMRS_0409_set2_v1.png}
    \caption{Torque-mixing magnetic resonance spectroscopy for a $z$-torque at 8.02 MHz. RF drives along $\hat{y}$ and $\hat{z}$ combine to reveal a magnetic resonance near 200 MHz in the low-bias field regime. Torques are measured at the difference frequency between the two drive fields which are swept to maintain constant offset between them in order to take advantage of signal amplification at mechanical resonances. \label{SuppFig:TMRS}}
    \label{fig:enter-label}
\end{figure}

Magnetic resonance near 208 MHz is measured via torque-mixing magnetic resonance spectroscopy (TMRS), as presented in \cite{Losby2015}.This technique is performed via simultaneous application of perpendicular drive fields. To excite a magnetic resonance, a sinusoidal drive along $H_y$ is applied with frequency $f_1$. An orthogonal drive field along $H_z$ is applied with frequency $f_2$ to produce RF torques at the mixed frequencies of the two drives, $f = f_1 \pm f_2$. The difference frequency is fixed at a low-frequency mechanical resonance to ensure signal enhancement for measurement of torques. The mechanical resonance used in the measurement presented in \autoref{SuppFig:TMRS} is a $z$-torque mode at 8.02 MHz (details in \cite{Hajisalem2019}). The TMRS spectrum shows a broad, field-dependent magnetic resonance around 208 MHz in low bias fields, providing external verification of an overlap between magnetic and mechanical resonances in \autoref{fig:magnres} of the main text.

% Existence of the gyrotropic mode at 208 MHz is confirmed through torque mixing magnetic resonance spectroscopy, following similar procedure as reported in \cite{Losby2015}. Perpendicular $H_x^{\mathrm{RF}}$ and $H_y^{\mathrm{RF}}$ drive fields are applied to the resonator such that the difference frequency between the two drives is constant at a mechanical resonance about the $z$-axis at 8.02 MHz (details of the 8.02 MHz resonance in \cite{Hajisalem2019}). The resulting torque signal from the mixed drive fields is presented in \autoref{SuppFig:TMRS}, showing a strongly field-dependent magnetic resonance around 200 MHz in low bias fields. Torque-mixing data was collected by measuring torque as a function of frequency at discrete field steps. 

\end{document}